\documentclass{article}

\usepackage{amssymb,amsfonts,amsmath}
\usepackage[dvips]{graphicx}

\usepackage{pstricks}
\usepackage{multirow}
\usepackage{cite}

\usepackage{lscape}
\usepackage{rotating}

\usepackage{setspace}
\usepackage[font={sf}]{caption}

\title{\Huge The biased evolution of generation time}

\author{M\'elissa Verin$^{1,2}$, Salom\'e Bourg$^1$, Fr\'ed\'eric Menu$^{1}$ and Etienne Rajon$^{1,*}$ \vspace{0.5cm}\\
\small $^{1}$CNRS, UMR5558, Laboratoire de Biom\'etrie et Biologie Evolutive, \\
\small Universit\'e de Lyon, F-69000, Lyon, Universit\'e Lyon 1, F-69622, Villeurbanne, France \vspace{0.3cm}\\
\small $^2$Section of Population Genetics, Center of Life and Food Sciences Weihenstephan\\
\small Technische Universit\"at M\"unchen, Freising, Germany \vspace{0.3cm}\\
\small $^*$Corresponding author: etienne.rajon@univ-lyon1.fr
}
\date{} 
\begin{document}

\maketitle

\begin{abstract}
\normalsize Many life-history traits, like the age at maturity or adult longevity, are important determinants of the generation time. For instance, semelparous species whose adults reproduce once and die have shorter generation times than iteroparous species that reproduce on several occasions. A shorter generation time ensures a higher growth rate in stable environments where resources are in excess, and is therefore a positively selected feature in this (rarely met) situation. In a stable and limiting environment, all combination of traits (or strategies) that produce the same number of viable offspring on average are strictly neutral even when their generation times differ. We first study the evolution of life-history strategies with different generation times in this context, and show that those with the longest generation time represent the most likely evolutionary outcomes. Indeed, strategies with longer generation times generate fewer mutants per time unit, which makes them less likely to be replaced within a given time period. This `turnover bias' inevitably exists and favors the evolution of strategies with long generation times. Its real impact, however, should depend on the strength and direction of other evolutionary forces; selection for short generation times, for instance, may oppose turnover bias. Likewise, the evolutionary outcome depends on the strength of such selection and population size, comparably to other biases acting on the occurrence of mutations.
\end{abstract}

\section*{Introduction}


Generation time is a major feature of organisms, whose evolution is usually considered through the prism of life-history strategies -- \textit{i.e.} combinations of traits that often impact generation time, like adult survival or the age of reproductive maturity. Why there exist populations or species with long generation times can be intriguing, as in a stable non-limiting environment the most evolutionary successful strategy is the one that grows most rapidly \cite{takada_1995, orr_2009}. Life-history strategies that reduce generation time -- ``fast strategies'' hereafter -- thus benefit from a selective advantage over ``slow strategies'' in this context. Whenever population growth is limited, however, the advantage to a fast strategy may vanish. In this context, the selective advantage of a mutant is given by the number of viable offspring it is expected to produce, also known as its net reproductive rate $R_0$ \cite{bulmer_1994, metz_etal_2008, metz_geritz_2016}: whether the mutant breeds shortly or not, or produces few or many litters, is not seen by selection. 

Many life-history traits influence both the generation time and the $R_0$, such that their evolutionary dynamics are often thought to be dominated by selection \cite{stearns_2000}. Nonetheless, heritable changes in life-history traits may not necessarily be filtered stringently by selection because negative relationships (trade-offs) exist between these traits and restrict their ability to vary independently \cite{houle_1998}
. Life-history strategies are indeed distributed along trade-offs whose shapes are often sufficient to predict which strategy, if any, is the most likely to evolve by selection.

The role of trade-off shapes is well illustrated by the case of reproduction strategies, which can be thought as distributed along a trade-off between the number of litters produced during an adult's lifetime (`longevity' hereafter) and the size of each litter \cite{charnov_schaffer_1973, schaffer_1974, pianka_1976, takada_1995}. A convex trade-off means that litter size decreases rapidly with longevity, such that semelparous individuals, who reproduce only once, produce a larger number of offspring in their adult life on average. Such a trade-off may appropriately describe the case of a species where adults have a fixed amount of resources to allocate to either reproduction or maintenance, such that increasing adult survival -- through higher investment in maintenance -- would be at the cost of a decreased lifetime fecundity. Because semelparous individuals produce more offspring faster, semelparity outcompetes any other reproduction strategy under this trade-off. Now consider the case where adults can acquire resources during their lifetime: this new source of energy may partly or fully compensate for the energy allocated to maintenance, and make the trade-off less convex or linear. The acquired energy may even overcompensate the investment in maintenance -- and make the trade-off concave -- such that iteroparous (long lived) strategies have higher $R_0$s than the semelparous. 

One can always find a trade-off shape that keeps the $R_0$ constant among strategies; for the case of reproductive strategies, this condition is met when litter size decreases linearly as adult survival increases -- the linear trade-off described above. This is an ideal setup to identify other players in the evolutionary dynamics of slow/fast strategies than the obvious selective advantage provided by a higher $R_0$; Bulmer \cite{bulmer_1985}, for instance, used it to demonstrate that environmental variability may select for iteroparity in a density-dependent environment. 

We first use this setup and study the impact of neutral processes on the evolutionary dynamics of two classes of life-history strategies associated with generation time: reproductive strategies and development duration. We confirm that the strategies considered, although they yield different generation times, are neutral in this situation: none is expected to increase or decrease in frequency, on average, in a pairwise competition. Next we investigate their neutral evolutionary dynamics and show that the most probable strategy at the evolutionary equilibrium is the slowest one, \textit{i.e.} maximum longevity or duration of development. This is due to a newly discovered evolutionary process that we called `turnover bias', whereby slow strategies are replaced after a longer time on average, so they are more likely to be observed. We also verify that turnover bias still modifies the evolutionary outcome when a specific strategy is selected. Turnover bias is most impacting when neutral processes like genetic drift dominate over selection, which occurs when selection is weak and/or when the effective population size is small \cite{kimura_1962}. This defines the conditions where turnover bias should be an important determinant of the evolution of life-history strategies. 

\section*{Results}

\subsection*{Strategies with different generation times and equal $R_0$ are neutral}

We first consider a monomorphic population living in a stable environment with limited resources. The population is constituted by $N_{J}$ juveniles and $N_{A}$ adults, whose dynamics can be described by the discrete system (\ref{eq:res0}):
\begin{equation}\label{eq:res0}
\left\{\begin{aligned}
	&N_{J}(t+1)=N_{A}(t) \times F \times (1-\alpha) \times d + N_{J}(t) \times \gamma\\
	&N_{A}(t+1)=N_{A}(t) \times \alpha + N_{J}(t) \times s_J \times (1-\gamma)
\end{aligned}\right. ,
\end{equation}
where $s_J$ is the survival of juveniles becoming adults, $F$ the lifetime fecundity, and $d$ a density-dependent egg (or newborn) survival. There are two ways of changing the generation time in this model: by increasing adult longevity -- by increasing the probability that an adult survives until the next reproductive season, $\alpha$ -- or by increasing the duration of development -- by increasing the probability that a juvenile remains a juvenile at the next reproductive season, $\gamma$. We study the evolution of $\alpha$ and $\gamma$ separately -- while setting the other, $\gamma$ or $\alpha$, to $0$ -- in the two distinct models (denoted 1 and 2 hereafter) described in fig. \ref{fig:models}. As a consequence, all strategies have the same age at maturity ($1$) when reproduction strategies evolve, and are all semelparous when development duration evolves.

\paragraph{Model 1.}The reproduction strategy is described by the continuous variable $\alpha$, the survival of adults from one reproductive event to the next. Adults with $\alpha=0$ are semelparous: they reproduce once and die. Iteroparity arises as soon as $\alpha$ is above $0$, and the mean number of reproductive events increases with $\alpha$ -- it equals $1/(1-\alpha)$. Fecundity per reproduction event equals $F(1-\alpha)$, which ensures that the number of offspring produced by an adult within its lifetime is independent of its reproduction strategy, so that all strategies have equal $R_0$.

Each egg or newborn produced survives with the density-dependent probability $d$, which decreases as the overall number produced, $N_{A}(t) F (1-\alpha)$, increases. We do not assign a specific density-dependence function to $d$ at this point; Instead, we assume that this function eventually yields a non-zero equilibrium where $N_{A}(t+1)=N_{A}(t)=N_{A}$ and $N_{J}(t+1)=N_{J}(t)=N_{J}$. This property is shared by many classical density-dependence functions, given a parameterization that avoids cyclic or chaotic dynamics (see SI texts 1-2 and figs. S1-S4). In consequence, the population dynamics simplify to:
\begin{equation}\label{eq:res1}
\left\{\begin{aligned}
	&N_{J}=N_{A} \times F \times (1-\alpha) \times d\\
	&N_{A}=N_{A} \times \alpha+ s_J \times N_{J}
\end{aligned}\right. ,
\end{equation}
for model 1 -- the system (\ref{eq:res1}) corresponds to (\ref{eq:res0}) with $\gamma=0$. By solving the system, we show that at equilibrium the product of the lifetime potential fecundity $F$ and of the density-dependent egg survival $d$ equals $1/s_J$. This property also emerges when $d$ is formulated explicitly (SI text 1 and figs. S1-S4). This implies that all strategies reach a stable equilibrium where they have the same lifetime density-dependent fecundity.

We now model the evolutionary dynamics of the reproduction strategy by considering the fate of a single mutant with strategy $\alpha_m$ appearing in a monomorphic population (with strategy $\alpha$) whose dynamics are described by the system (\ref{eq:res1}). We use here the adaptive dynamics framework, which commonly assumes that mutations are rare enough for the resident population to reach an equilibrium before a mutant appears \cite{metz_etal_1992, geritz_etal_1998}. As we have seen, at this equilibrium the resident's lifetime fecundity $F \times d$ equals $1/s_J$. Another classical assumption of adaptive dynamics is that the resident population is large enough for the mutant to be negligibly rare at the beginning of the invasion. The density-dependent parameter $d$ thus only depends on $N_{A}(t) F (1-\alpha)$ and the mutant's lifetime fecundity $F \times d$ equals that of the resident, $1/s_J$ -- we relax the two assumptions above later (see \textit{Turnover bias in polymorphic populations}), with no visible impact on our results. The dynamics of the mutant can thus be modeled with the following system:
\begin{equation}\label{eq:mut1}
\left\{\begin{aligned}
	&N_{Jm}(t+1)=N_{Am}(t) \times (1-\alpha_m) / s_J\\
	&N_{Am}(t+1)=N_{Am}(t) \times \alpha_m + s_J \times N_{Jm}(t)
\end{aligned}\right.
\end{equation}
where $(1-\alpha_m)/s_J$ is the number of surviving eggs laid by the mutant at time $t$. In the appendix, we show that all mutants have equal growth rates -- regardless of the resident they compete with -- such that no reproduction strategy should increase (or decrease) in frequency in response to selection. Reproduction strategies are therefore selectively neutral, or \textit{iso-neutral} in the terminology of Proulx and Adler \cite{proulx_adler_2010}. 

\paragraph{Model 2.}The generation time can also be impacted by the duration of development, which increases as $\gamma$ increases in equation (\ref{eq:res0}). With $\alpha=0$, the dynamics of the mutant -- with strategy $\gamma_m$ -- is described by the following system:
\begin{equation}\label{eq:mut2}
\left\{\begin{aligned}
	&N_{Jm}(t+1)=N_{Am}(t) \times 1 / s_J + N_{Jm}(t) \times \gamma_m\\
	&N_{Am}(t+1)=s_J \times N_{Jm}(t) \times (1-\gamma_m)
\end{aligned}\right.
\end{equation}
Here again, we find that the mutant's growth rate is insensitive to the mutant and the resident strategies (see appendix). This confirms that strategies with equal $R_0$s but different generation times are neutral. In the following section, we study their neutral evolutionary dynamics.

\subsection*{The neutral evolution of slow/fast strategies}

When changes in genotype frequencies are not -- or little -- governed by selection, neutral processes like genetic drift may affect evolution in finite populations. Here we study the neutral evolutionary dynamics of strategies with different generation times using a Markov chain that includes the production of offspring, and the occurrence and fixation of mutations impacting these strategies. 

\paragraph{Reproduction strategies.} As before, the number of offspring produced by an adult with reproduction strategy $\alpha_i$ at any time step equals $(1-\alpha_i)/s_J$, and each survives with probability $s_J$. With probability $\mu$, an offspring carries a mutation that changes its reproduction strategy. Here again, we assume that $\mu$ is small enough for a mutation to appear and fix (or go extinct) before another mutation occurs. Because reproduction strategies are neutral, a mutation fixes with probability $\epsilon$ regardless of the strategy it yields -- typically $\epsilon=1/N_e$, with $N_e$ the effective population size. Ignoring the transient dynamics of mutant fixation or loss, a population will always be monomorphic in this framework. 

We further assume that mutations produce a small change in the reproduction strategy, such that a population with strategy $\alpha_i$ can only mutate to the immediately lower $\alpha_{i-1}$ or higher $\alpha_{i+1}$. The probability of transition to either of these states equals:
\begin{align}\label{eq:mark}
	p_{i,i-1} = p_{i, i+1} &= \frac{1-\alpha_i}{s_J} \times s_J \times \frac{\mu}{2}\times \epsilon \\
									&= (1-\alpha_i) \times \frac{\mu}{2}\times \epsilon,
\end{align}
and the probability of remaining in state $i$ equals:
\begin{equation}\label{eq:mark2}
	p_{i,i} = 1 - p_{i, i+1} - p_{i, i-1}
\end{equation}
For the first and last of the $n_s$ states in the Markov chain, only the probabilities of transition $p_{1,2}$ and $p_{n_s, n_s-1}$ can be calculated with equation (\ref{eq:mark}), and the probabilities of remaining in those states are $p_{1,1} = 1 - p_{1, 2}$ and $p_{n_s,n_s} = 1 - p_{n_s,n_s-1}$, respectively.

The evolutionary dynamics of the reproduction strategy can thus be described by the Markov chain represented in figure~\ref{fig:mark}. Using equations (\ref{eq:mark}) and (\ref{eq:mark2}), one can create a Markov chain of any length by dividing the range of possible reproduction strategies into a given number of adjacent values of $\alpha$. We used the set $\{0, 0.01, ..., 0.99\}$ and calculated the equilibrium state of the Markov chain, given by the normalized eigenvector for the eigenvalue $1$ of the transition matrix defined by equations (\ref{eq:mark} - \ref{eq:mark2}), which we obtained with the \textit{markovchain} package of the R software \cite{markovchain,R_2013}.

At equilibrium, the most likely reproduction strategy has the highest level of iteroparity ($\alpha = 0.99$; Fig.~\ref{fig:mark_res}). Semelparity (\textit{i.e.} $\alpha=0$) is, among all possible reproductive strategies, the least likely to be observed: for example it would be observed less than $0.2 \%$ of the time on a very long time series where many neutral mutations would have fixed, or in less than $2$ populations in $1000$ at equilibrium. It should be noted that interpreting the equilibrium probability of a specific strategy is irrelevant if one does not know the set of strategies a population can really access: the probability of observing one specific strategy will necessarily decrease as the model includes more strategies, for instance. Only ranges of strategies are relevant, such as $\alpha \in [0,0.1]$, which would include about $1 \%$ of observations. Medium to high degrees of iteroparity -- \textit{i.e.} $\alpha > 0.5$, or more than $2$ reproductive events on average -- occur about $86 \%$ of the time. 

\paragraph{Turnover bias.} Slow strategies -- \textit{e.g.} high degrees of iteroparity -- are more likely to evolve neutrally as a result of a bias that we call `turnover bias'. For instance, a population with reproduction strategy $\alpha_{i-1}$ produces more offspring per reproduction event, and thus more mutants, than a population with a slightly higher level of iteroparity $\alpha_{i+1}$ -- this is known as the generation time effect \cite{thomas_etal_2010}. Therefore, a population that switches from state $i$ to state $i+1$ is less likely to switch back to $i$ than a population switching from $i$ to $i-1$ (see fig. \ref{fig:mark}). This yields a small bias towards increasing the degree of iteroparity at each step, which does not stop until the highest degree of iteroparity is reached. In other words, a population where a slow strategy has fixed generates fewer mutants per time unit and thus awaits a longer time before a different strategy reaches fixation. It is, therefore, more likely to be observed.

\paragraph{Duration of development.} The neutral evolutionary dynamics of reproduction strategies is easily understood when considering how the Adults class is filled at each timestep: a proportion $\alpha$ remain through adult survival, while a proportion $1-\alpha$ are newly made through reproduction. Only the latter mutate (but see the \textit{Age-dependent mutation rate} section), which explains the term $(1-\alpha_i)$ in equation \ref{eq:mark}. When treating the question of the evolution of development time, it is more intuitive to ask how the Juveniles class is filled. A proportion $\gamma$ are juveniles remaining juveniles, while a proportion $(1-\gamma) \times s_J$ become adults that will produce new juveniles at a mean rate $1/s_J$ -- thus a proportion $1-\gamma$ is generated this way. Of course, only the latter are subject to mutation, such that the evolutionary dynamics of $\gamma$ may be described by equations (\ref{eq:mark}--\ref{eq:mark2}), substituting $\alpha$ with $\gamma$. Turnover bias is therefore also expected to favor the evolution of long development durations, so long as these do not impede the $R_0$.

\subsection*{Turnover bias in polymorphic populations}

The models described in the two sections above rely on the assumption that mutations are rare, so that the population is only transiently polymorphic. Moreover, in the neutral Markov chain, mutants can only reach neighboring reproduction strategies, even though the degree of iteroparity or development duration might vary continuously. These assumptions can be relaxed in individual-based simulations where mutations occur with probability $\mu$ and change the offspring strategy from its parent's by an amount sampled from a continuous distribution (see \textit{Simulation procedure} in the Material and methods section). The model otherwise matches the discrete life cycle described above (see equation~\ref{eq:res0}). 

At each timestep, a juvenile with genotype $i$ can either remain a juvenile (with probability $\gamma_i$) or attempt to become an adult (probability $1-\gamma_i$); then it survives with probability $s_J$. An adult with genotype $i$ survives with probability $\alpha_i$ at each time step and produces $N_{Ei} = F (1-\alpha_i)$ eggs on average. The $\sum_{i} N_{Ei}$ eggs produced by all genotypes are in competition for survival. The egg survival $d$ needs to be defined explicitly here: we use the exponential $e^{-\sum_{i} N_{Ei}/K}$ throughout this paper, such that survival is close to $1$ when $\sum_{i} N_{Ei}$ is small and decreases as this number increases. As in models 1 and 2 above, we model the evolution of $\alpha$ and $\gamma$ separately, while fixing $\gamma_i$ or $\alpha_i$, respectively, to $0$ for all $i$. 

\begin{figure}[h]
\centerline{\includegraphics[height=150mm, angle=270]{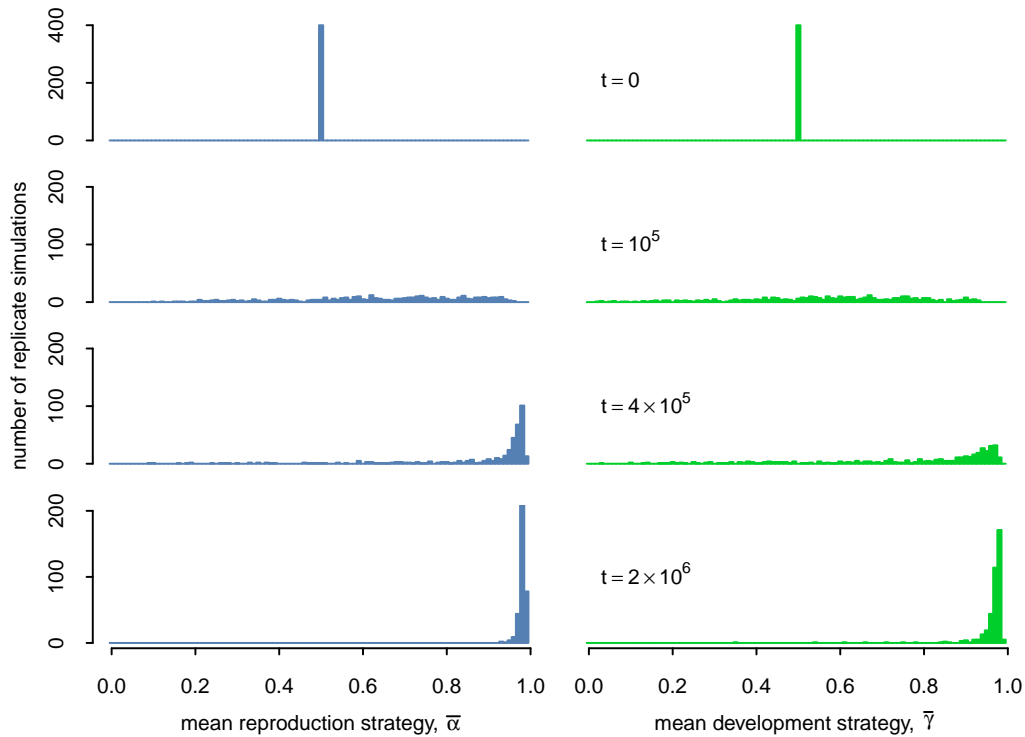}}
\caption{Simulations confirm that slow turnover, exemplified by a long adult lifetime -- \textit{i.e.} a high $\alpha$ in model 1 -- and a long development time -- a high $\gamma$ in model 2 -- should evolve in polymorphic populations living in a stable density-dependent environment. The simulation procedure is described in the Material and methods section. The distribution of the mean values of $\alpha$, $\overline{\alpha}$, is represented at different simulation times for $400$ replicates of model 1 (left panels); right panels similarly represent the distribution of the mean values of $\gamma$ evolving in model 2. Parameter values: $F=5$, $\mu=0.001$, $K=1000$.}
\label{fig:res_sim}
\end{figure}

We ran $400$ replicate simulations of the evolution of $\alpha$ and $\gamma$ with $\mu=0.001$, $F=5$ and $K=1000$. The simulations were initiated with a single genotype with $\alpha_1=0.5$ or $\gamma_1=0.5$ (fig.~\ref{fig:res_sim}, $t=0$). Both traits have very similar evolutionary dynamics: initially, the replicate populations diverge and exhibit a large range of strategies (fig.~\ref{fig:res_sim}, $t=10^5$), with the means of the distributions of $\overline \alpha$ and $\overline \gamma$ above $0.5$. At higher simulation times (fig.~\ref{fig:res_sim}, $t=4 \times 10^5$ and $t=2 \times 10^6$), the mean strategy is, in most replicate simulations, very close to its maximum possible value -- that is, to the strategy with the longest generation time. 

Genotypes with a slow turnover evolve under a wide range of parameter values, including even higher mutation rates, lower fecundity and different initial values of $\alpha_1$ or $\gamma_1$ (see SI text 2 and fig. S5). We also studied another form of density-dependency and obtained very similar results to those presented in figure~\ref{fig:res_sim} (SI text 2 and fig. S6). 

It is worth noting that the population size is not strictly constant in this model: the density dependent process can yield population sizes above or below its expected equilibrium. This may impact the resulting evolutionary dynamics, because fast strategies perform better when the whole population growth rate is above $1$, while slow strategies are more able to wait for better conditions when this rate is below $1$. These time-varying competitive advantages might not compensate each others exactly, which could give a selective advantage to a specific strategy. We thus ran simulations of the evolution of the reproduction strategies ($\alpha$), this time with a strictly constant population size of $1000$. The evolutionary dynamics in these simulations are very similar to those presented in fig.~\ref{fig:res_sim} (SI text 2 and fig. S7). These results confirm that slow strategies are generally expected to evolve in a stable, density-dependent environment.

\subsection*{Age-dependent mutation rate}

Above we considered a constant mutation rate among the various litters produced by an individual. By doing so, we neglected the well-known fact that, in many species, the mutation rate can increase with one or both parents' age \cite{risch_etal_1987, crow_2000, petit_hampe_2006, kong_etal_2012}. This increase is likely due to a large extent to an increase in the number of germline cell divisions \cite{hurst_ellegren_1998}. This raises two distinct issues: that iteroparous individuals may be biased towards producing more mutants later in life, and that maybe this yields an increase in the overall mutation rate as the degree of iteroparity increases. The first issue likely remains a minor one as long as the overall mutation rate is small: iteroparous individuals will produce mutants late in life instead of throughout, which will not affect the evolutionary dynamics as long as the overall mutation rate remains equal between strategies.

The second issue is more problematic and needs to be resolved here: an increase in the mutation rate with the degree of iteroparity could, in theory, revert the neutral dynamics and make semelparity the most likely outcome under the linear trade-off. It is trivial, from equations (\ref{eq:mark}) and (\ref{eq:mark2}), that a mutation rate $\mu$ increasing from a basal rate $\mu_0$ at a higher rate than the function $\mu=\mu_0 / (1-\alpha_i)$ would produce such a result, so we need to address the question `how much does the mutation rate per generation increase with iteroparity?'. We need to build a model to obtain this relationship, which as far as we know has not been determined empirically. 

We assume that a batch of gametes is produced each time an individual reproduces, of a size proportional to the number of offspring it is expected to yield at this reproductive season (SI text 3). The overall number of gametes produced is thus constant among strategies, yet our model shows that producing them at once -- as semelparous do -- reduces the overall number of cell divisions in the germline. In consequence, the average mutation rate per gamete increases with longevity, but the difference is negligible whenever the number of gametes produced in a lifetime is large. For instance, if $10^6$ gametes should be produced in a lifetime, an iteroparous individual with $\alpha=0.9$ -- living $10$ reproductive seasons on average -- would have an average mutation rate increased by $10.97 \%$ compared to a semelparous individual. This number raises to $23.18 \%$ if the number of gametes produced in a lifetime equals $1000$.   

How may this phenomenon affect the neutral evolution of reproduction strategies? In the previous sections, we showed that highly iteroparous strategies were scarcely replaced due to their long generation times. A higher mean mutation rate of iteroparous individuals might (over)compensate the bias and favor the evolution of semelparity. We replaced the mutation rate in the Markov chain described by equations (\ref{eq:mark}) and (\ref{eq:mark2}) with that given by equation (\ref{eq:mean}), such that the equilibrium now depends on the number of gametes produced in a lifetime, $n_g$. We consider that the number of batches $n_{bi}$ of a strategy with adult survival $\alpha_i$ equals its mean longevity, $1/(1-\alpha_i))$. We set the mutation rate per cell division to $10^{-5}$ and checked that this parameter has no impact on the equilibrium distribution. In figure~\ref{fig:mark_res2}, we show the resulting Markov chain equilibrium with $n_g=10^6$: the highest degree of iteroparity remains the most likely evolutionary outcome, although in this situation semelparity and weak iteroparity are more likely to evolve than when aging is ignored. Decreasing $n_g$ to the unrealistically low value of $100$ does not change these results qualitatively (fig.~\ref{fig:mark_res3}).

\subsection*{Turnover bias \textit{vs.} selection}

It is legitimate to ask whether turnover bias can have an impact on the evolutionary dynamics of generation time in the presence of selection. Consider first \textbf{a two-allele haploid model} where the two alleles yield the two reproduction strategies $\alpha_1$ and $\alpha_2$ ($\alpha_1<\alpha_2$). Turnover bias favors the evolution of $\alpha_2$ because this strategy generates fewer mutations per time unit, and selection can counteract this process by increasing the fixation probability of $\alpha_1$ (denoted $\epsilon_1$) and decreasing that of $\alpha_2$ ($\epsilon_2$) from their value in the neutral model ($\epsilon$ in the previous section). The bias is entirely canceled when:
\begin{equation}\label{eq:eq}
	(1-\alpha_1) \epsilon_1 = (1- \alpha_2) \epsilon_2,
\end{equation}
such that it is equally probable to evolve to $\alpha_1$ from a population where $\alpha_2$ is fixed than the reverse. Assuming that the fitness advantage of $\alpha_1$ over $\alpha_2$ equals $s$ (and the fitness disadvantage of $\alpha_2$ $-s$), we can calculate \cite{rajon_plotkin_2013}:
\begin{equation}
	\epsilon_1=\dfrac{1-e^{-2 s}}{1-e^{-2 N s}} \text{ and } \epsilon_2=\dfrac{1-e^{2 s}}{1-e^{2 N s}}. 
\end{equation}
The equilibrium in equation \ref{eq:eq} is thus obtained for:
\begin{equation}
	s=\dfrac{\log(1-\alpha_1) - \log(1-\alpha_2)}{2 (N-1)}. 
\end{equation}
$s$ increases as the difference between $\alpha_1$ and $\alpha_2$ increases -- \textit{i.e.} when turnover bias is more acute -- and when the population size decreases and selection consequently becomes less efficient. For instance, in a population of size $1000$, a semelparous allele ($\alpha_1=0$) would need to provide a selective advantage of about $0.0012$ to be as likely to evolve as an iteroparous allele with $\alpha_2=0.9$.

Now consider, as we did in previous sections, that reproduction strategies can \textbf{vary continuously}, distributed along a trade-off between litter size and longevity. The shape of the trade-off determines the selective value of any strategy competing with others. A convex trade-off appropriately sets a selection gradient in the opposite direction to turnover bias. We model such a trade-off by changing the number of eggs produced by an adult with genotype $i$ in our simulations, $N_{Ei}$, from $F(1-\alpha_i)$ to $F(1-\alpha_i)^S$, with $S \geq 1$. The trade-off is linear when $S=1$ and becomes increasingly convex as $S$ increases, making selection for semelparity stronger. 

\begin{figure}[h]
\centerline{\includegraphics[height=87mm, angle=270]{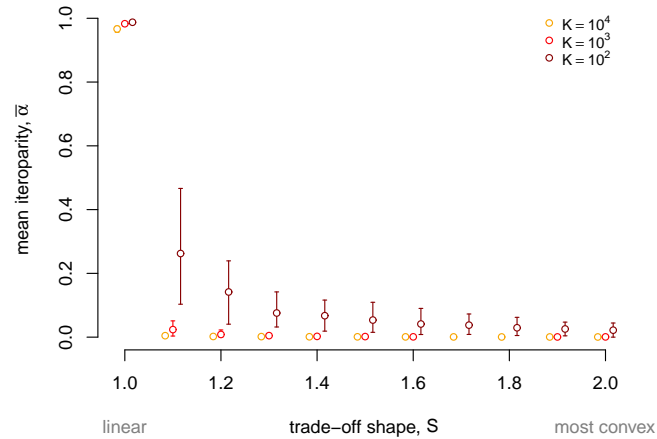}}
\caption{Both the strength of selection and population size are key parameters in the evolution of reproduction strategies. Population size increases with parameter $K$ in our model, which sets the level of resources in the density-dependent egg survival function. The trade-off between litter size and adult survival becomes more convex as $S$ increases, which increases the selection pressure for semelparity (\textit{i.e} for $\alpha=0$). Accordingly in our simulated populations, high levels of iteroparity evolve when $S=1$ -- the linear trade-off -- but semelparity evolves as $S$ increases. The evolutionarily expected level of iteroparity decreases with $S$ at a speed that increases with $K$.  For each set of parameters, the mean (circle) and the range from the $10 \%$- to the $90 \%$-quantile are represented.}
\label{fig:selection}
\end{figure}

The evolutionarily expected reproduction strategies are represented in fig.~\ref{fig:selection}. In large populations ($K=10^4$), semelparity evolves at very low values of $S$ -- as soon as the trade-off becomes non-linear in fig.~\ref{fig:selection}. In smaller populations, the level of iteroparity $\alpha$ decreases with $S$, at a slower rate when the population size is smaller ($K=10^2$ \textit{vs} $K=10^3$). This result is typical of evolutionary dynamics where neutral and selective processes act in opposite directions: selection is most efficient in large populations -- favoring the evolution of semelparity in the present case -- such that  turnover bias has a small effect on the evolutionary dynamics in this situation. As the population size decreases, selection becomes less efficient and the evolving reproduction strategy is the result of a balance between turnover bias and selection. 

\section*{Discussion}

Life-history strategies that influence the generation time but not the lifetime reproductive success are known to be neutral in a density-dependent environment yielding stable population dynamics \cite{bulmer_1994, takada_1995}, suggesting a minor role of the generation time in their evolution. Here we show that a bias favors the evolution of the slowest possible life-history strategy -- that with the longest generation time. Indeed, individuals adopting (fast) strategies with short generation times produce many offspring per time unit (and thus many mutants), while slow strategies instead favor the survival of their carriers (who do not mutate). Therefore, a fast strategy is more likely to be replaced shortly, and is less likely to be observed than a slow strategy. This logic should be applicable to any trait or strategy associated with generation time. We modeled the evolution of two of them: reproduction strategies, where the fast semelparous compete with the slow iteroparous, and developmental timing where juveniles can develop readily (fast) or wait (slow).

Turnover bias -- as we called it -- is an unescapable part of the evolutionary dynamics of strategies with different generation times. This bias impacts the occurrence of mutations but not their fixation; we actually detected it by building an ``origin-fixation model'' (OFM) that separates the two processes. OFMs are rarely applied to life-history strategies, which might partly explain why turnover bias has not been identified earlier -- the framework of adaptive dynamics is often preferred, which takes its roots in an OFM to build the canonical equation but then usually ignores differences in the mutation-generating process among alternative strategies \cite{dieckmann_law_1996, champagnat_etal_2001, mccandlish_stoltzfus_2014}. To our knowledge, only Proulx and Day \cite{proulx_day_2001} have built an OFM to study the evolution of a trait associated with the generation time. They have considered differences in fixation probabilities among strategies, but not in their rates of mutation appearance per time unit, rendering it impossible to detect the bias. Interestingly, OFMs are ideal but not necessary to identify turnover bias: we still observe the bias in simple individual-based simulations (see fig.~\ref{fig:res_sim}) -- where origin and fixation are not formally separated -- which are rather common in theoretical evolutionary ecology.

An expected feature of biases of the mutational process is that their contribution to evolutionary dynamics is maximum when alternative alleles have similar fixation probabilities, that is when allele fixation is mainly due to neutral processes like genetic drift \cite{stoltzfus_2012, mccandlish_stoltzfus_2014}. Our results indeed indicate that turnover bias impacts the evolutionary outcome more strongly when the selective values of the strategies in competition are close and/or when the population is small. When selection strongly favors the fixation of one specific strategy (or of a range thereof), turnover bias is still acting but its effects are negligible. We should thus expect to observe turnover bias after the fixation of some neutral or nearly neutral mutations, which may take a long time compared to the fixation of advantageous mutations. Turnover bias may therefore only yield differences along evolutionary timescales where numerous neutral substitutions occur. It might thus explain a part of the extraordinary diversity of life-history strategies reported across large phylogenetic trees (\textit{e.g.} \cite{stearns_1983, gaillard_etal_1989, Salguero-Gomez_etal_2016}), while its role at the population level may be contingent on specific population parameters. 

The importance of turnover bias in nature thus relies on how often selection is inefficient among strategies with different turnovers. It might be a very special and rare situation, but the factual truth is that we lack empirical evidence to support this claim: selection among strategies distributed along a trade-off depends on the shape of the trade-off, which is scarcely inferred. This is likely due to statistical issues: relatively small samples allow the detection of negative relationships between traits, whereas inferring precise relationships between traits requires many accurate observations. A promising approach consists in obtaining trade-off shapes through mechanistic models of the genetic, physiological and ecological processes that contribute to them \cite{braendle_etal_2011} -- evolutionary epidemiologists have studied the virulence-persistence trade-off using mechanistic models \cite{alizon_van-baalen_2005}, but we are not aware of their application to other trade-offs. 

We have ignored other, potentially important selection pressures that may contribute to the evolutionary dynamics of strategies associated with the generation time. Bet-hedging strategies, for instance, evolve as a response to environmental fluctuations and are often characterized by longer generation times than those selected by a constant environment, as exemplified by iteroparity or facultative dormancy \cite{cohen_1966, bulmer_1985, seger_brockmann_1987, orzack_tuljapurkar_1989, rajon_etal_2009, rajon_etal_2014}. This prediction contrast with a recent study by Mitteldorf and Martins \cite{mitteldorf_martins_2014} showing that a fluctuating environment can select for short generation times, by creating repeated episodes of directional selection. This gives a selective advantage to genotypes adapting quickly, which can be achieved by fast strategies as they produce more mutants per time unit. This apparent theoretical conflict -- a fluctuating environment can favor both slow and fast strategies -- needs to be resolved. In any case, the role of turnover bias in this context remains unclear, as the fluctuating environment both provides a selection pressure for a specific strategy and reduces drastically the effective population size. We have also ignored the role of deleterious mutations, which are known to be the most frequent kind of mutations in a stable environment \cite{eyre_keightley_2007}. Purging selection in presence of these mutations would likely favor genotypes producing fewer mutants. This might create a selection pressure for slow strategies, even in stable environments. 

We have not considered the genetic architecture of the evolving traits, as it cannot change turnover bias directly. Turnover bias results from traits impacting the generation time, regardless of their genetic, physiological or developmental components. Nonetheless, the genetic architecture may play a (distinct) major role in the evolutionary dynamics of the traits considered, as it should determine the mutation rates from one strategy to others. If theses rates are uneven, a mutational bias should occur that may change the evolutionary outcome. Mutation biases are known to play a role in the evolution of molecular traits \cite{bulmer_1991, rokyta_etal_2005, stoltzfus_yampolsky_2009, shah_gilchrist_2011} but they are rarely considered for phenotypic traits (but see \cite{xue_etal_2015}). Mutation biases are similar to turnover bias in the sense that they affect the origin of mutations but not their fixation.

Overall, evolutionary ecology typicallly focused on the complex interactions of organisms with their biotic and abiotic environment and the selection pressure that results, leaving the study of neutral processes and biases in the production of mutants to theoretical population genetics. Our hybrid model makes novel predictions, which in our opinion praises for more dialogue between these two fields.


\section*{Material and methods}

	\subsection*{Simulation procedure} 
	We simulated a population of a variable number $n$ of genotypes with different reproduction strategies $\alpha_i$. Each genotype $i$ is represented by $N_{Ji}$ juveniles and $N_{Ai}$ adults. At a given timestep, adults with genotype $i$ produce a number of eggs $N_{Ei}$ sampled from a Poisson distribution with mean $N_{Ai} \times F \times (1-\alpha_i)$. Eggs survive with the density dependent probability $e^{-\sum_{i=1}^N N_{Ei}/1000}$; for each genotype, the number of surviving eggs is sampled from a binomial with this probability and $N_{Ei}$ trials. Note that we also use a different density-dependent function in the supplement. The surviving eggs constitute the juveniles at the beginning of the next timestep.\\
	Each offspring of genotype $i$ can mutate with probability $\mu$. When this event occurs, the number of genotypes $n$ is incremented by $1$ and the new genotype $n$ has $N_{An}=0$, $N_{Jn}=1$, and $\alpha_n=\alpha_i +\epsilon$. $\epsilon$ is sampled from a normal distribution with mean $0$ and standard deviation $0.05$; mutants with $\alpha_N$ below $0$ or above $0.99$ take the value $0$ and $0.99$, respectively. The number of juveniles with genotype $i$ at the beginning of the next timestep is decremented by $1$.\\
	After reproduction, the number of surviving adults of genotype $j$ is sampled from a binomial with $N_{Aj}$ trials and probability $\alpha_j$. These will constitute the adults with genotype $i$ at the next timestep, together with the offspring produced at the previous timestep that survive, whose number is sampled from a binomial with $N_{Ji}$ trials and probability $s_J$. At each timestep, a genotype is suppressed if $N_{Ai}=N_{Ji}=0$. \\
	In each replicate simulation, we simulated the process described above during $2 \times 10^6$ timesteps, and recorded $N_{Ai}$, $N_{Ji}$ and $\alpha_i$ for each of the $i$ genotypes present in the population. We ran $100$ replicate simulations for each parameter set explored; the distribution of the mean values of $\alpha$ in the $100$ replicates is represented in fig. \ref{fig:res_sim} at different timesteps. The program was written in the R programming language and is available on demand. 

\section*{Appendix: mutant's asymptotic growth rate}
Initially, the mutant is a juvenile so $N_{Jm}(0)=1$ and $N_{Am}(0)=0$. The mutant population may grow from this point, and its ability to do so is given by the Lyapunov exponent, or invasion fitness, denoted $\log \lambda (\alpha_m, \alpha)$ \cite{metz_etal_1992, vandooren_metz_1998, rand_etal_1994, ferriere_gatto_1995, roff_2008}:
\begin{equation}\label{eq:lyapunov}
	\log \lambda (\alpha_m, \alpha) = \displaystyle \lim\limits_{T \to \infty} \frac{1}{T} \log \frac{N(T)}{N(0)},
\end{equation}
where $N(t)=N_{Jm}(t) + N_{Am}(t)$. 

Because $F \times d$ is equal for all residents, the invasion dynamics of the mutant are independent of the resident's strategy -- either $\alpha$ for the model 1 or $\gamma$ for model 2 (see system [\ref{eq:mut1}]). For model 1, the asymptotic growth rate of the mutant can be calculated as the first eigenvalue of the matrix:
\begin{equation}\label{eq:mat}
M_1=\begin{pmatrix}
	\alpha_m & s_J \\
	(1-\alpha_m) / s_J & 0
\end{pmatrix}.
\end{equation}

The mutant's asymptotic growth rate is calculated as the largest eigenvalue of $M_1$ (defined in eq. [\ref{eq:mat}]), which can be obtained by solving $|M_1-\lambda I|=0$. This yields the characteristic polynomial:
\begin{equation}
	(\alpha_m -\lambda) \times (- \lambda) - (1-\alpha_m)/s_J \times s_J =0,
\end{equation}
 which simplifies to:
\begin{equation}\label{eq:fact}
	(\lambda-1) (\lambda +1 - \alpha_m)=0
\end{equation}
Hence the polynomial has roots $\lambda_1=1$ and $\lambda_2=\alpha_m-1$. $\lambda_1>\lambda_2$ for $0 \leq \alpha_m < 1$ so the asymptotic growth rate of the mutant equals $1$.

The transition matrix for model 2 can be written:
\begin{equation}\label{eq:mat2}
M_2=\begin{pmatrix}
	0 & (1-\gamma_m) \times s_J \\
	1 / s_J & \gamma_m
\end{pmatrix}. 
\end{equation}

Solving $|M_2-\lambda I|=0$, we obtain the characteristic polynomial:
\begin{align}
	(\gamma_m -\lambda) \times (- \lambda) - (1-\gamma_m)/s_J \times s_J &=0\\
	\Leftrightarrow \; (\lambda-1)(\lambda + 1 - \gamma_m) &=0,
\end{align}
which mirrors equation (\ref{eq:fact}) above: here, too, the mutant's asymptotic growth rate equals $1$, regardless of its strategy.


\clearpage


\section*{Figures}

\psset{xunit=0.7cm,yunit=0.7cm}
\begin{figure}[h]
\begin{center}
	\begin{pspicture}(-4, 0.5)(4, -8.5)
		\rput[l](-4.5,0){\footnotesize \textbf{\textsf{Model 1}}}
		\pscircle(-2.5,-2){5mm}	
		\rput[c](-2.5,-2){\textsf J}					
		\pscircle(2.5,-2){5mm}	
		\rput[c](2.5,-2){\textsf A}						
		\pscurve[linewidth=1pt, arrowsize=5pt]{->}(-1.5,-1.5)(0,-1)(1.5, -1.5)
		\rput[c](0, -0.5){\footnotesize $\mathsf{s_J}$}
		\pscurve[linewidth=1pt, arrowsize=5pt]{<-}(-1.5,-2.5)(0,-3)(1.5, -2.5)
		\rput[c](0, -3.5){\footnotesize $\mathsf{F \times (1- \alpha) \times d}$}
		\pscurve[linewidth=1pt, arrowsize=5pt]{->}(3.5,-1.5)(4,-2)(3.5, -2.5)
		\rput[l](4.25, -2){\footnotesize $\mathsf{\alpha}$}
		\psframe(-5,0.5)(5, -4.25)
		\psframe(-5, -4.25)(5,-9)

		\rput[l](-4.5,-5){\footnotesize \textbf{\textsf{Model 2}}}
		\pscircle(-2.5,-7){5mm}	
		\rput[c](-2.5,-7){\textsf J}					
		\pscircle(2.5,-7){5mm}	
		\rput[c](2.5,-7){\textsf A}						
		\pscurve[linewidth=1pt, arrowsize=5pt]{->}(-1.5,-6.5)(0,-6)(1.5, -6.5)
		\rput[c](0, -5.5){\footnotesize $\mathsf{(1-\gamma) \times s_J}$}
		\pscurve[linewidth=1pt, arrowsize=5pt]{<-}(-1.5,-7.5)(0,-8)(1.5, -7.5)
		\rput[c](0, -8.5){\footnotesize $\mathsf{F \times d}$}
		\pscurve[linewidth=1pt, arrowsize=5pt]{<-}(-3.5,-6.5)(-4,-7)(-3.5, -7.5)
		\rput[r](-4.25, -7){\footnotesize $\mathsf{\gamma}$}
	\end{pspicture}
\end{center}
\caption{The two models considered can produce different population turnovers -- by varying $\alpha$ in model 1 and $\gamma$ in model 2 -- without affecting the lifetime reproductive success. Increasing $\alpha$ in model 1 increases the average longevity of adults (A), and increasing $\gamma$ in model 2 increases the average time before juveniles (J) become adults. Both models are special cases of the more general model described by the system (\ref{eq:res0}); model 1 is obtained by setting $\gamma=0$ and model 2 is obtained by setting $\alpha=0$.}
\label{fig:models}
\end{figure}
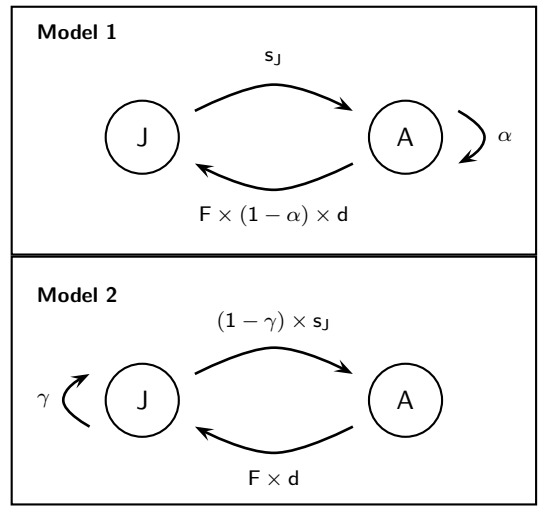

\psset{xunit=1cm,yunit=1cm}
\begin{figure}[h]
\begin{center}
	\begin{pspicture}(-4, -1.3)(4, -3)
		\pscircle(-2.5,-2){6mm}	
		\rput[c](-2.5,-2){\textsf{i-1}}					
		\pscircle(0,-2){6mm}	
		\rput[c](0,-2){\textsf i}					
		\pscircle(2.5,-2){6mm}	
		\rput[c](2.5,-2){\textsf{i+1}}		
		\psline[linewidth=3pt, arrowlength=1]{->}(-1.7,-1.8)(-0.8,-1.8)
		\rput[c](-1.25, -1.5){\footnotesize $\mathsf{p_{i-1,i}}$}
		\psline[linewidth=2pt, arrowlength=1]{<-}(-1.7,-2.2)(-0.8,-2.2)
		\rput[c](-1.25, -2.5){\footnotesize $\mathsf{p_{i,i-1}}$}
		\psline[linewidth=2pt, arrowlength=1]{->}(0.8,-1.8)(1.7,-1.8)
		\rput[c](1.25, -1.5){\footnotesize $\mathsf{p_{i,i+1}}$}
		\psline[linewidth=1pt, arrowlength=1]{<-}(0.8,-2.2)(1.7,-2.2)
		\rput[c](1.25, -2.5){\footnotesize $\mathsf{p_{i+1,i}}$}
		\psline[linestyle=dotted]{->}(-4.2,-1.8)(-3.3,-1.8)
		\psline[linestyle=dotted]{<-}(-4.2,-2.2)(-3.3,-2.2)
		\psline[linestyle=dotted]{->}(3.3,-1.8)(4.2,-1.8)
		\psline[linestyle=dotted]{<-}(3.3,-2.2)(4.2,-2.2)
	\end{pspicture}
\end{center}
\caption{Schematic representation of the model for the neutral evolution of the reproduction strategy. The states $i-1$, $i$ and $i+1$ correspond to populations with different reproduction strategies, with $\alpha_{i-1}<\alpha_i<\alpha_{i+1}$ -- \textit{i.e.} the degree of iteroparity increases from left to right. The values of $p_{i, i-1}$, $p_{i-1, i}$, $p_{i+1, i}$ and $p_{i, i+1}$ can be calculated using eq. [\ref{eq:mark}]; higher probabilities are represented by thicker arrows. $p_{i, i-1}$ equals $p_{i, i+1}$ according to eq. (\ref{eq:mark}), is lower than $p_{i-1, i}$ (because $1-\alpha_{i-1}>1-\alpha_{i}$) and is higher than $p_{i+1, i}$ (because $1-\alpha_{i+1}<1-\alpha_{i}$). The probabilities of remaining in a given state are not shown, but can be calculated from eq. (\ref{eq:mark2}).}
\label{fig:mark}
\end{figure}
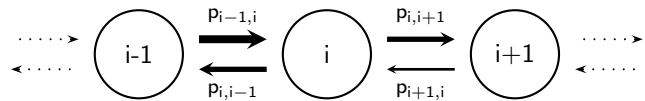

\begin{figure}[h]
\centerline{\includegraphics[height=87mm, angle=270]{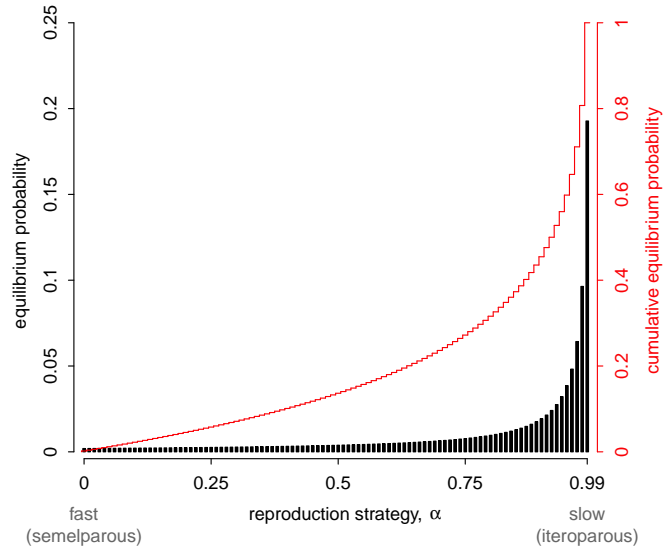}}
\caption{High degrees of iteroparity are the most likely evolutionary outcome for a regulated population living in stable environment. In this context, only neutral processes can impact the evolution of the reproduction strategy. The black bars (left \textit{y}-axis) represent the equilibrium probability that a given population will have a given $\alpha$, under the neutral model described in Fig.~\ref{fig:mark}. The red line (right \textit{y}-axis) represents the probability that a population be at a given $\alpha$ or below.}
\label{fig:mark_res}
\end{figure}

\begin{figure}[h]
\centerline{\includegraphics[height=87mm, angle=270]{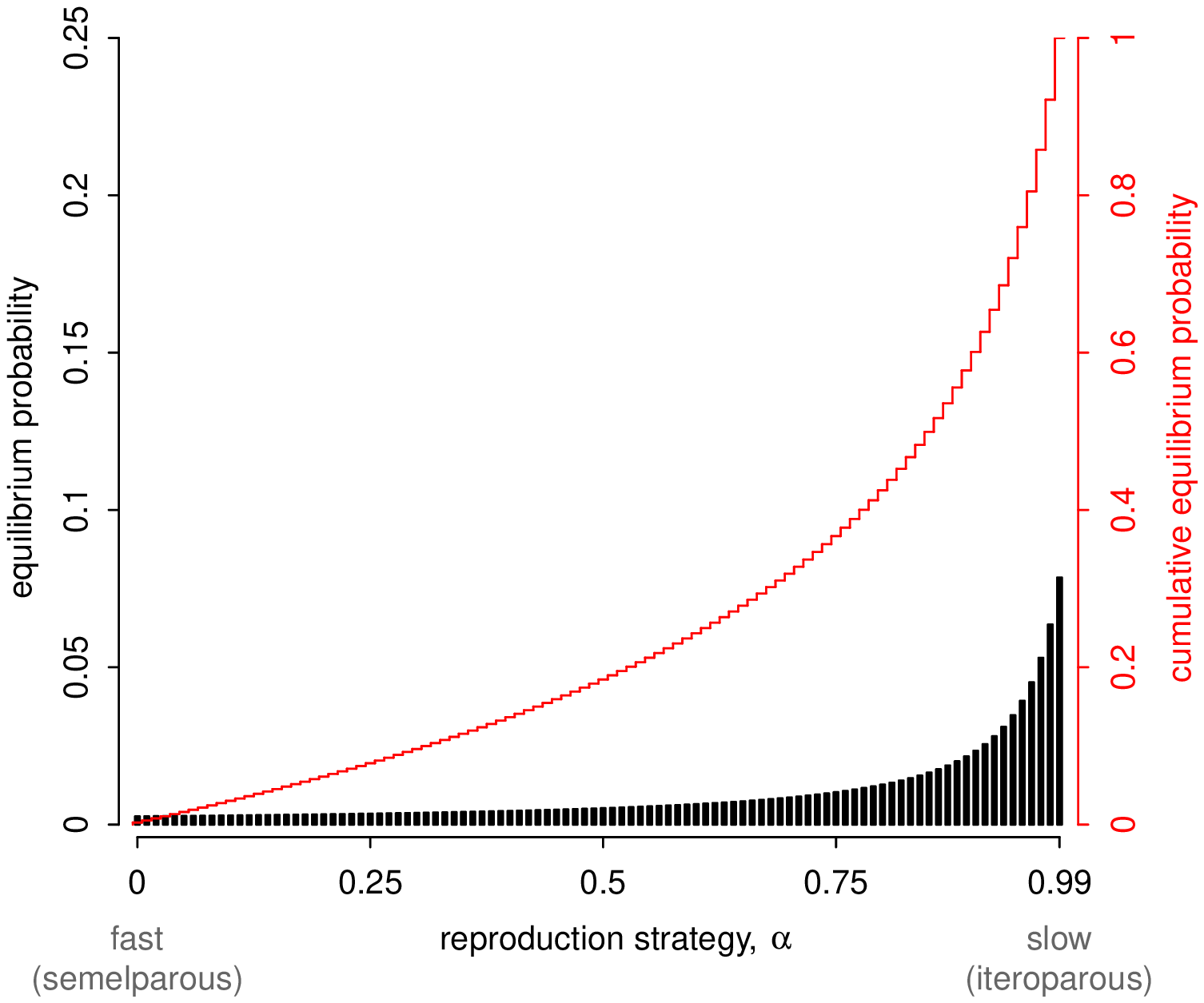}}
\caption{A high degree of iteroparity remains the most likely evolutionary outcome when the mutation rate can increase with an individual's age. The figure is similar to fig.~\ref{fig:mark_res}, except that the model includes the increase in the mutation rate with $\alpha$ (see text). The number of gametes produced in a lifetime, $n_g$, equals $10^6$.}
\label{fig:mark_res2}
\end{figure}

\clearpage
\section*{Supplementary information}

\renewcommand{\thesubsection}{SI~text~\arabic{subsection}:}
\renewcommand{\thefigure}{S\arabic{figure}}
\renewcommand{\thetable}{S\arabic{table}}
\renewcommand{\theequation}{S\arabic{equation}}
\setcounter{figure}{0}

\subsection*{SI text 1: Density-dependent population dynamics}

Here we study the population dynamics described by the system [1] in the main paper, replacing $d$ by two density-dependency functions.

\subsubsection*{Density-dependency function 1: $d=e^{-\sum_i{N_{Ei}/K}}$}

\begin{figure}[!h]
\centering \includegraphics[height=160mm, angle=270]{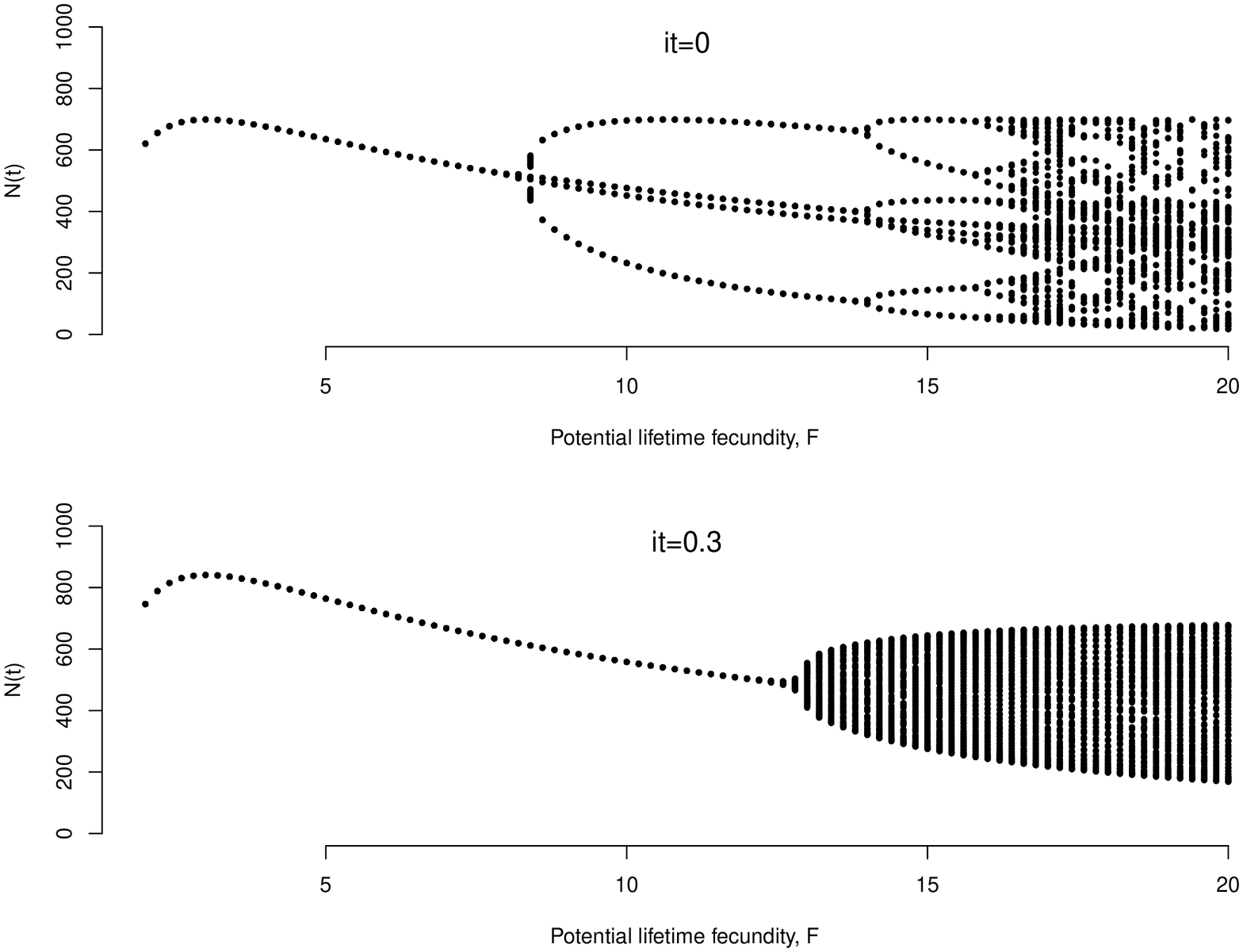}
\caption{Bifurcation diagram for the density-dependency function 1. For each value of $F$, the total population size ($N_A+N_J$) is represented for $100$ timesteps after an initial simulation period of $100$ steps. More than one value of $N(t)$ mean that the equilibrium is unstable (cyclic or chaotic), which occurs here when $F>8$ (top panel) or $F>13$ (bottom). }
\label{fig:bif1}
\end{figure}

Here the population is monomorphic, so the number of eggs produced at a given timestep, $\sum_i N_{Ei}$ equals $N_{A}(t) \times F \times (1-\alpha)$. We simulated the system with $s_J=0.9$ and $F=5$, starting with $N_A=10$ and $N_J=10$. As shown in fig.~\ref{fig:bif1}, this function yields a stable dynamic equilibrium when $F<8$, with juvenile survival $s_J=0.9$ and adult survival $\alpha=0$ (\textit{i.e} semelparous). Lower $s_J$ yield unstable population dynamics at higher values of $F$ (\textit{e.g.} at $F=15$ for $s_J=0.5$). Increasing $it$ has a similar impact, as shown in fig.~\ref{fig:bif1} (bottom panel) for $\alpha=0.3$.

\begin{figure}
\centering 
\includegraphics[height=160mm, angle=270]{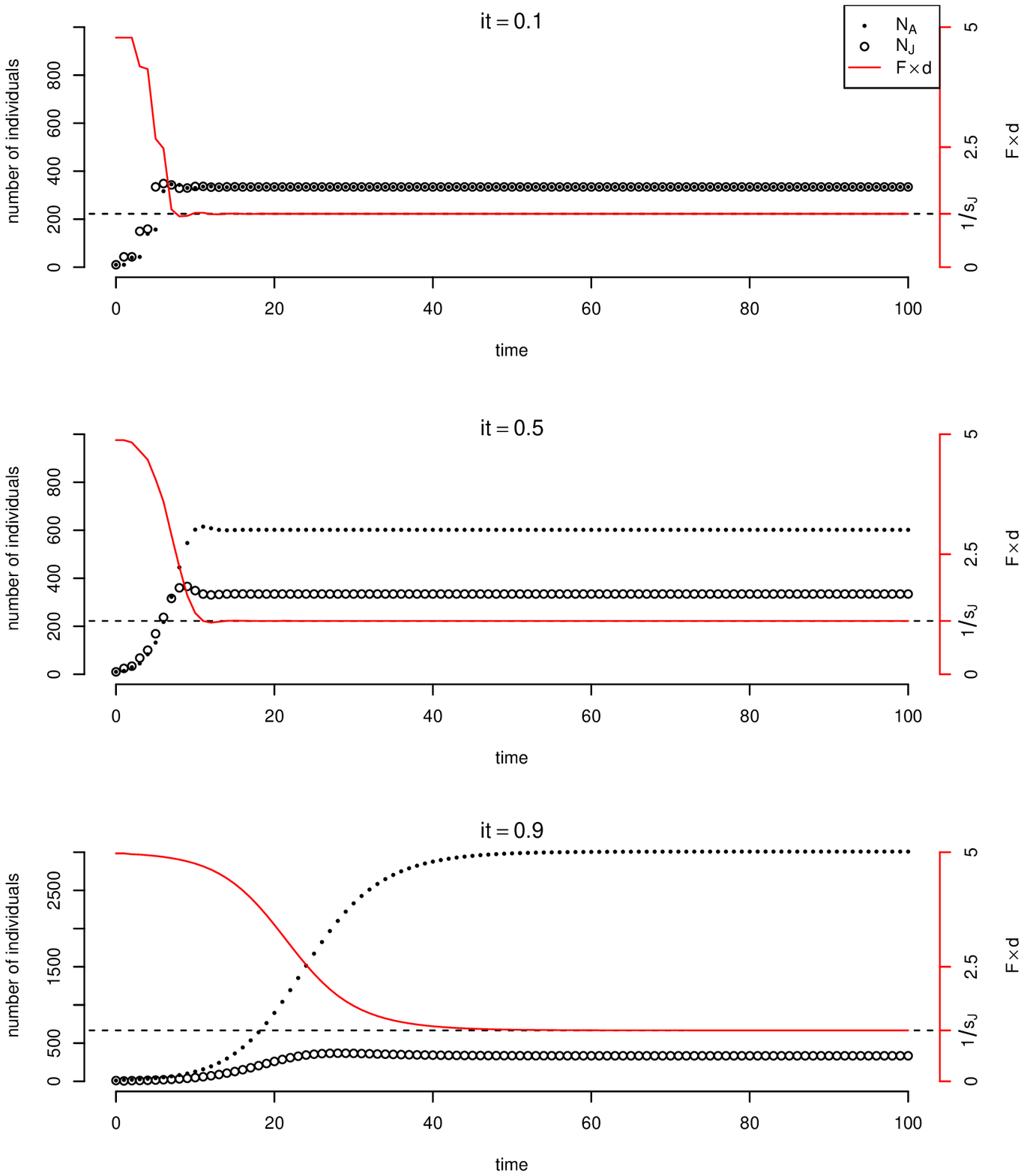}
\caption{Population dynamics simulated using the density-dependency function 1. }
\label{fig:dyn1}
\end{figure}

We ran simulations in the stable regime ($F=5$, $s_J=0.9$) for $100$ generations with $\alpha=0.1$, $0.5$ and $0.9$ (fig. \ref{fig:dyn1}). The population equilibrium is indeed stable and quickly reached. At the equilibrium, the density-dependent lifetime fecundity $F \times d$ equals $1/s_J$, as expected from the more general theoretical approach in the main paper. 

\subsubsection*{Density-dependency function 2: $d=\dfrac{1}{1+e^{\beta \times (\sum_i{N_{Ei}}-K)}}$}

\begin{figure}
\centering \includegraphics[height=160mm, angle=270]{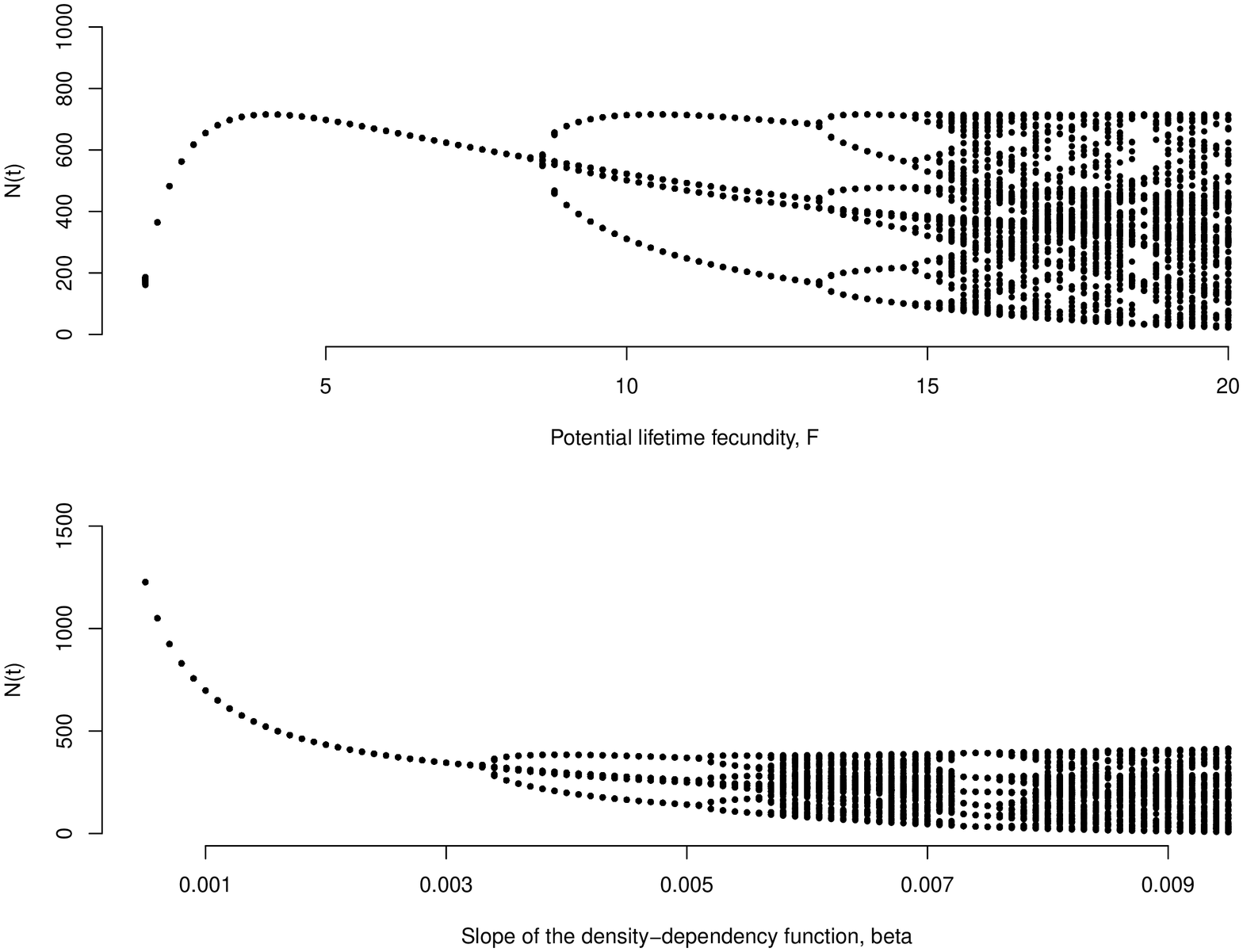}
\caption{Bifurcation diagram for the density-dependency function 2. For each value of $F$ (top panel) or $\beta$ (bottom), the total population size ($N_A+N_J$) is represented for $100$ timesteps after an initial simulation period of $100$ steps. More than one value of $N(t)$ mean that the equilibrium is unstable (cyclic or chaotic), which occurs here when $F>8$ (top panel) or $\beta>0.0033$ (bottom). }
\label{fig:bif2}
\end{figure}

This function yields a sigmoidal relationship between $d$ and the number of eggs produced, $\sum_i{N_{Ei}}$ -- as before, the population is monomorphic at this point so $\sum_i N_{Ei}=N_{A}(t) \times F \times (1-\alpha)$. Two parameters control the shape of the function: $K$ is the number of eggs at which $d=0.5$, and $beta$ controls the slope of the function at this point (it is steeper at higher $\beta$). As before, the population dynamics become cyclic or chaotic as $F$ increases; For the parameters used in fig.~\ref{fig:bif2} (top panel, $\beta=0.001$, $K=400$, $s_J=0.9$, $\alpha=0$), this occurs when $F>8$. The shape of the function plays an obvious role, as shown in the bottom panel of fig.~\ref{fig:bif2}, where we see that the population dynamics are unstable when $\beta$ is over $0.0033$. 

\begin{figure}
\centering 
\includegraphics[height=160mm, angle=270]{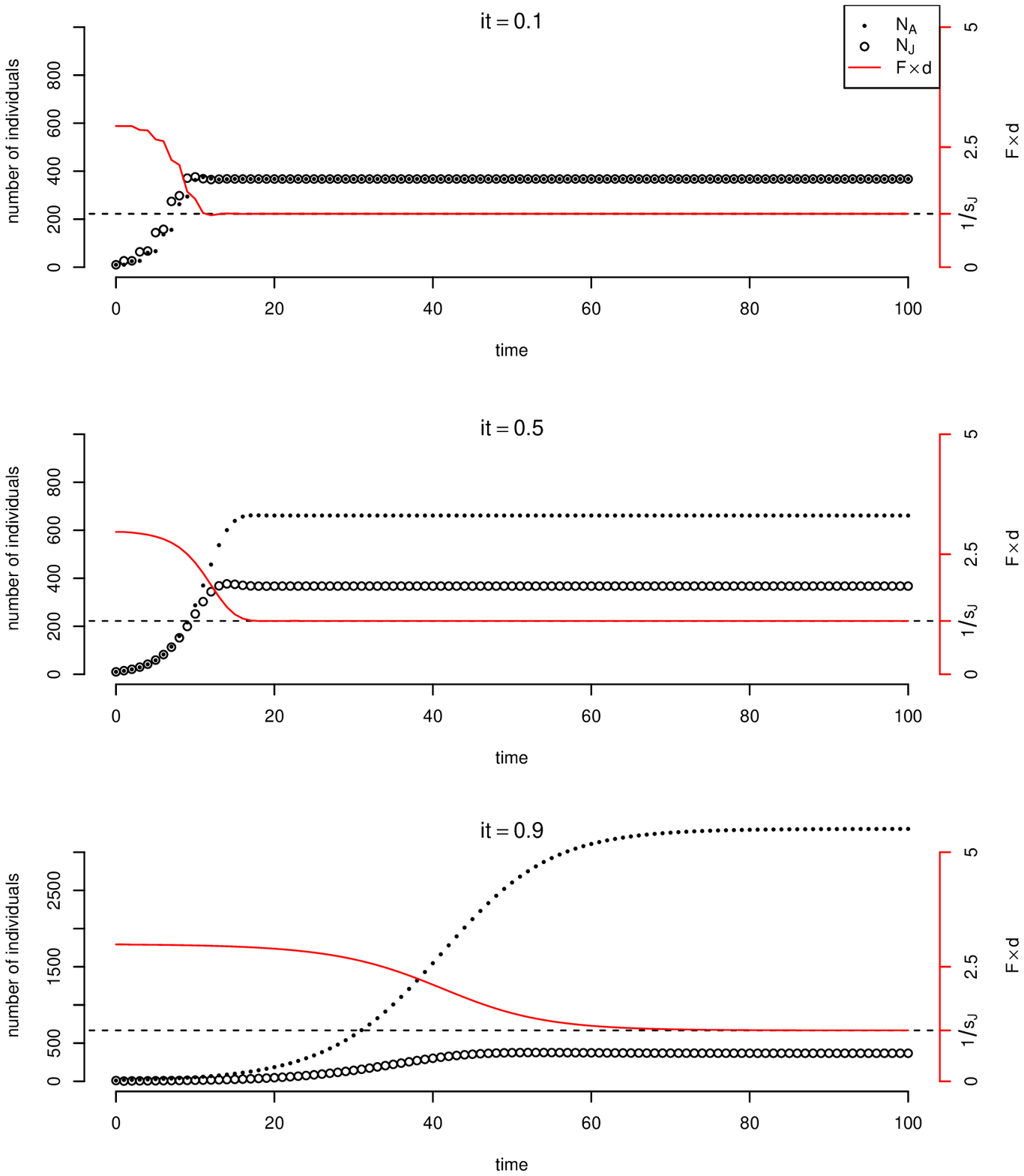}
\caption{Population dynamics simulated using the density-dependency function 2. }
\label{fig:dyn2}
\end{figure}

We also ran simulations in the stable regime with this function ($F=5$, $s_J=0.9$, $\beta=0.001$) for $100$ generations with $\alpha=0.1$, $0.5$ and $0.9$ (fig. \ref{fig:dyn2}). At the equilibrium, the density-dependent lifetime fecundity $F \times d$ equals $1/s_J$, as with the first function above and as we show should generally be expected in regulated populations with stable population dynamics (see main text). 

\subsection*{SI text 2: Evolutionary dynamics}
	 	
\subsubsection*{Density-dependency function 1: $d=e^{-\sum_i{N_{Ei}/K}}$}

\begin{figure}
\centering 
\includegraphics[height=160mm, angle=270]{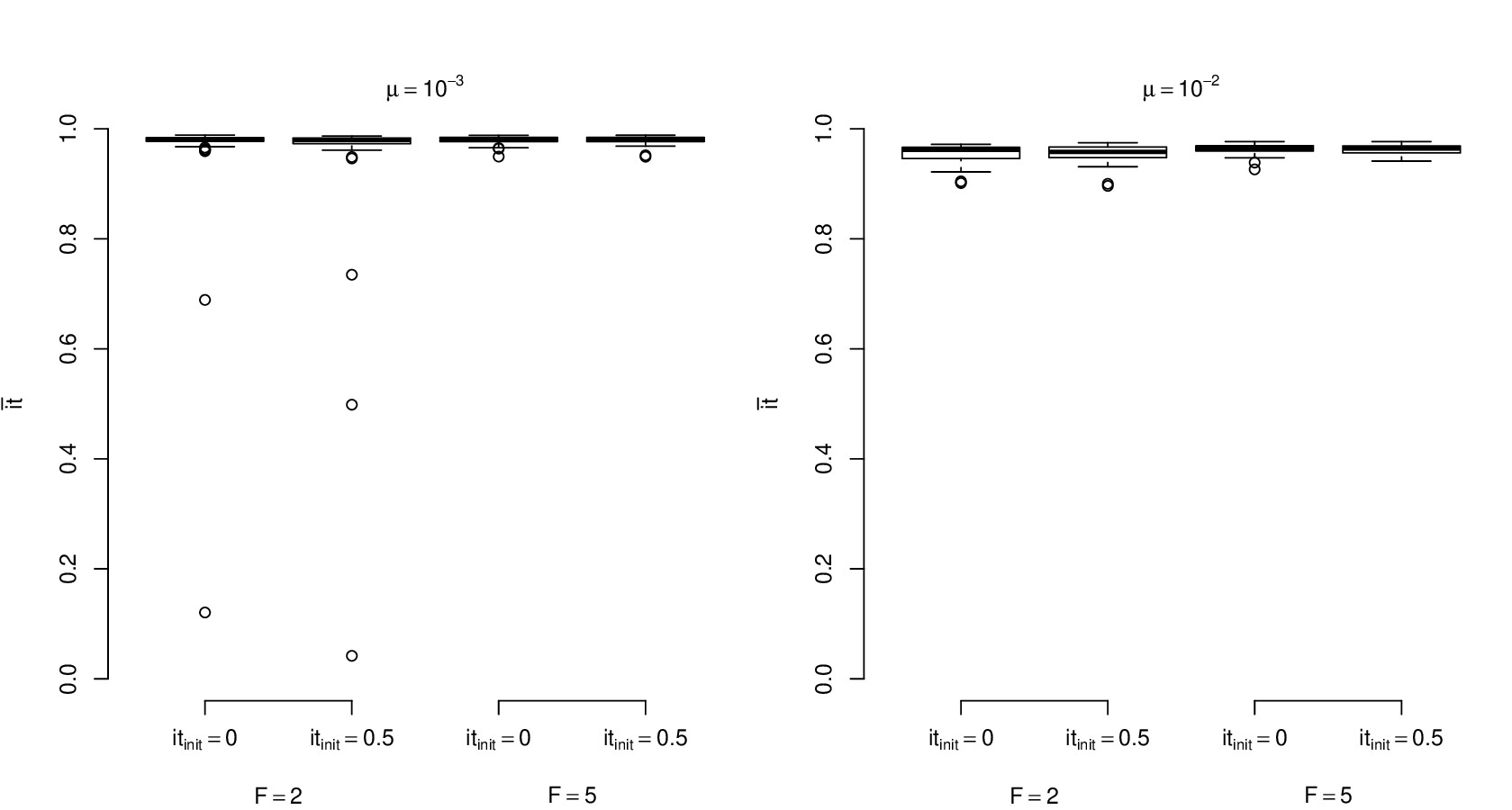}
\caption{Distribution of the mean value of $\alpha$ evolving in $100$ replicate simulations (for each parameter set) where the density dependent egg survival is modeled by function 1, with $K=1000$. We used two high mutation rates $\mu=10^{-3}$ (left panel) or $\mu=10^{-2}$ (right), with two values for the potential fecundity $F$ ($2$ or $5$). We initiated the simulations with two initial values of $\alpha$: $0$ and $0.5$. }
\label{fig:evol1}
\end{figure}

We simulated the evolution of $it$ in $100$ replicate populations, as described in the Material and methods section of the main paper. We used different parameter sets, showing that neither the mutations rate $\mu$, the potential fecundity $F$ nor the initial value of $\alpha$ in the population change the evolutionary outcome (fig.~\ref{fig:evol1}): in every case studied, a very high $\alpha$ evolves, presumably as a result of neutral evolutionary dynamics (see paper). Note that the simulations were run for $2 \times 10^6$ generations when $\mu=10^{-3}$ and for $10^{-5}$ generations when $\mu=10^{-2}$. 

\subsubsection*{Density-dependency function 2: $d=\dfrac{1}{1+e^{\beta \times (\sum_i{N_{Ei}}-K)}}$}

In this section, the function for the density-dependent is replaced by $d=\dfrac{1}{1+e^{\beta \times (\sum_i{N_{Ei}}-K)}}$. Otherwise the simulation procedure is exactly the same as that described in the Material and methods section in the main text and in the previous section. Changing $d$ does not impact the results: in most replicate simulations, a very high level of iteroparity evolves, independently of the mutation rate $\mu$, potential fecundity $F$ and initial level of iteroparity $\alpha$.

\subsubsection*{Constant population size}

Our aim here is to maintain the size of the population strictly constant to a value $N$. We sample the number of surviving adults of each genotype (subscript $i$) from a binomial with probability $\alpha_i$, and the number of juveniles with genotype $i$ that become adults from a binomial with probability $s_J$. The eggs that survive and become juveniles at the next timestep equals $N$ minus the sum of surviving adults and juveniles. These are sampled according to their frequency in the pool of eggs, \textit{i.e.} an egg will be of genotype $i$ with probability: 
\begin{equation*}
p_i=\dfrac{N_{Ai} \times F \times (1-\alpha_i)}{\displaystyle \sum_{j=1}^{n_{genotypes}} N_{Aj} \times F \times (1-\alpha_j)}
\end{equation*}

\begin{figure}[h]
\centerline{\includegraphics[height=87mm, angle=270]{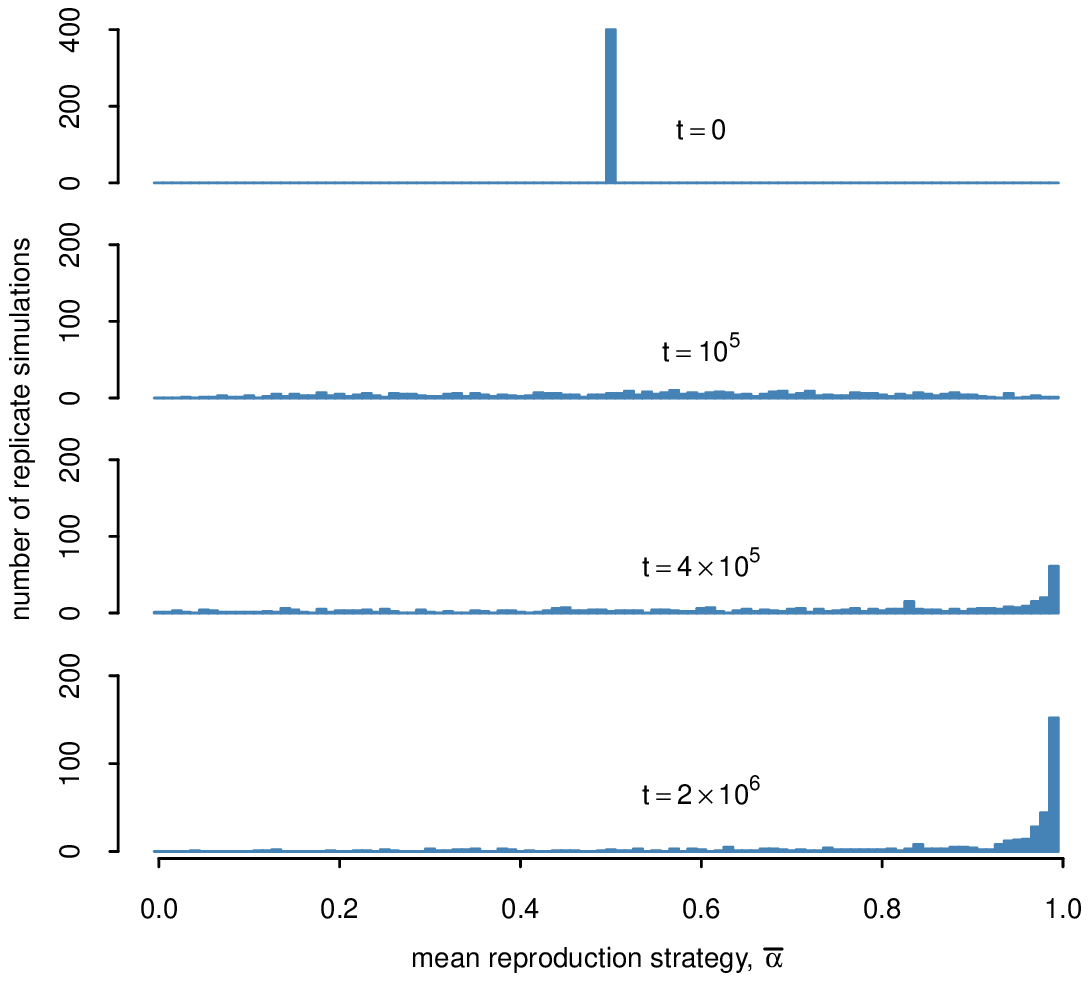}}
\caption{Slow turnover reproduction strategies also evolve when the population is kept strictly stable (see SI text 2). The distribution of the mean values of $\alpha$, $\overline{\alpha}$, is represented at different simulation times for $400$ replicates of model 1. Parameter values: $F=5$, $\mu=0.001$.}
\label{fig:Nconst}
\end{figure}

The evolutionary dynamics obtained with a constant population size are show in fig.~\ref{fig:Nconst}.

\subsection*{SI text 3: Age-dependent mutation rate}

In this section, we detail the calculations of the age-specific mutation rate, and of the resulting average mutation rate per gamete, when gametes need to be produced on several occasions in a lifetime. Since gametes are typically short-lived, we will consider that a new batch of gametes needs to be produced at each reproductive season. Thus the average number of batches $n_b=1/(1-\alpha)$. Semelparous will produce $1$ batch, iteroparous with $\alpha=0.75$ will produce $4$, etc. Moreover, we assume that the number of gametes to produce in a batch is proportional to the number of offspring produced at the corresponding reproductive season. Since all strategies have equal lifetime fecundities, the overall number of gametes, $n_g$, is constant across strategies. 

\begin{sidewaysfigure}[h]
\centerline{\includegraphics[height=210mm, angle=270]{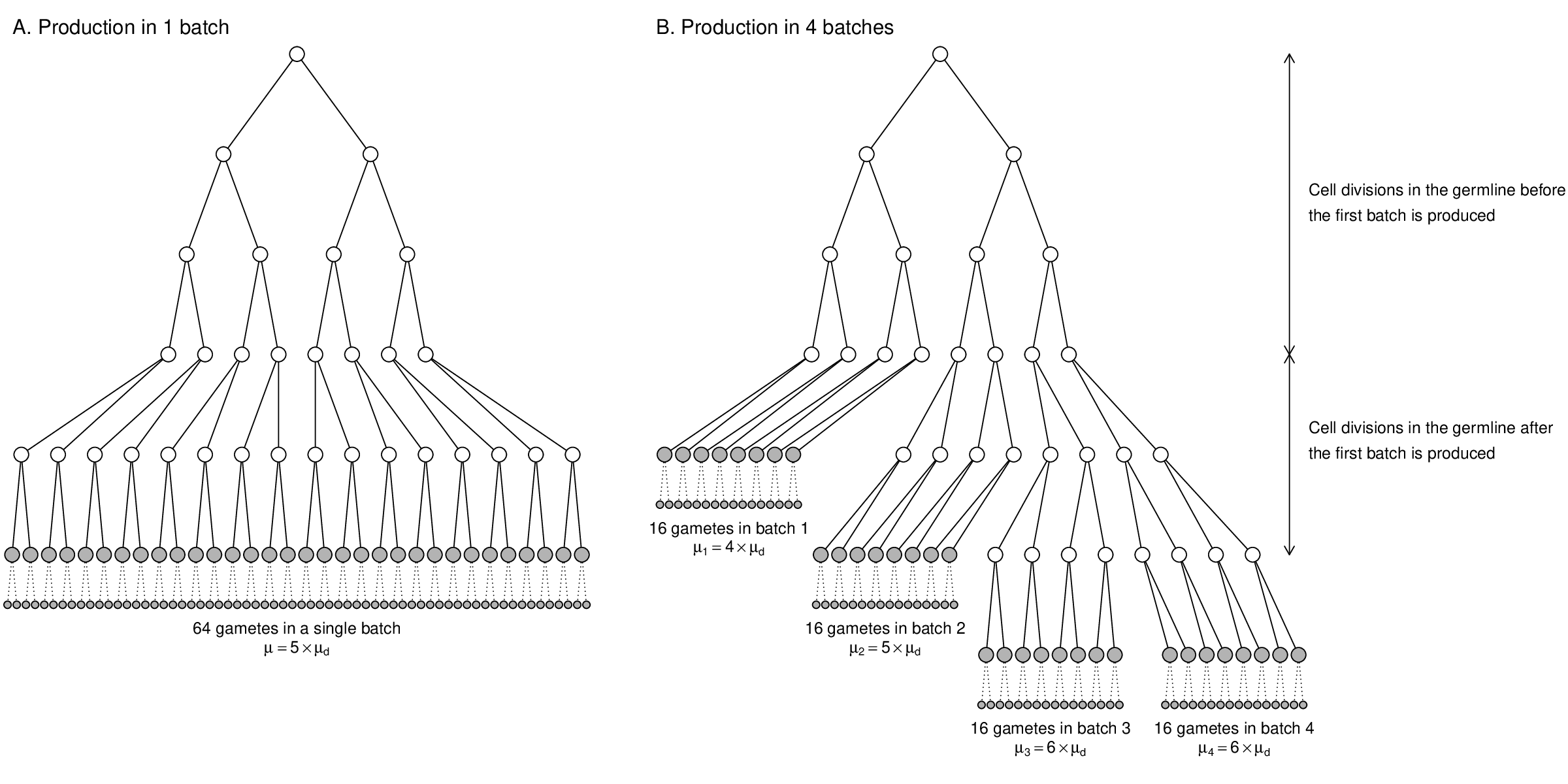}}
\caption{A depiction of the two production strategies considered. Germline cells ongoing mitosis are represented as white circles, while cells and gametes produced through meiosis are represented in grey. Cell divisions involving DNA replication (and mutation) are represented by solid lines, whereas the second division of meiosis, during which DNA is not replicated, is represented by dotted lines. The vertical position of the gametes shows their mutation rate, which is equal for all gametes produced in a single batch. Gametes produced in several batches also have an equal mutation rate within a batch, but more cell divisions are needed to produce gametes in old batches, so the mutation rate increases with age. The mutation rate in the first batch of four is lower than that of gametes produced in a single batch (this is generally true when the number of batches is above $2$), and higher in the two last batches. On average, continuous production yields a higher mutation rate whenever the number of batches $n_b \geq 3$ (see SI text 3).}
\label{fig:gametes}
\end{sidewaysfigure}

A critical number of cells in the germline needs to be reached before the first gametes can be produced through meiosis. Here we assume that half the germline cells are used to produce a batch of gametes, the others being kept for future production -- this is true in man, for which each primordial (Ad) spermatogonia divides into another Ad cell and a differentiated Ap cell that will eventually yield spermatozoids \cite{clermont_1966}. Thus the number of germline cells needs to reach a critical threshold equal to twice the number of cells required for a batch (see fig.~\ref{fig:gametes}). Then each germline cell destined to immediate production yields $4$ gametes through meiosis. Thus, from the number of gametes in a batch (${n_g}/{n_b}$), one can obtain the number of germline cells -- and thus the number of cell divisions -- that needs to be reached before meiosis starts (fig.~\ref{fig:gametes}). This number equals 
\begin{equation}
t_0=\log_2 \left( 2 \times \dfrac{n_g}{n_b} / 4 \right) =\log_2 \left( \dfrac{n_g}{n_b} \right)-1,
\end{equation}
where $\dfrac{n_g}{n_b} / 4$ is the number of germline cells used to produce the first batch of gametes. Substituting for $n_b$, we obtain $t_0=\log_2 \left(n_g \times (1-\alpha) \right)-1$.

\paragraph*{Age-distribution of the mutation rate.}

The first batch is produced from cells mutated $t_0$ times. They will experience another division with DNA replication -- and mutation -- during meiosis I, and another division without DNA replication (meiosis II), so the mutation rate of batch $1$, $\mu_1$, equals $t_0 + 1$ times the mutation rate per cell division, $\mu_d$.\\
Then one division needs to occur each time another batch is produced, such that the mutation rate of batch $i$ ($i \in \{1, .., n_b -1 \}$) equals
\begin{equation}\label{eq:mut_batch}
\mu_i=\mu_1+ (i-1) \times \mu_d= (t_0+1) \times \mu_d +(i-1) \times \mu_d = (t_0 + i) \times \mu_d = (\log_2(\dfrac{n_g}{n_b}) + i -1) \times \mu_d
\end{equation}
The last batch has the same mutation rate as the previous one because the remaining germline cells are used, with no further DNA replication required, so that $\mu_{n_b} = \mu_{n_b-1}$.

\paragraph*{Mean mutation rate.}

Equation (\ref{eq:mut_batch}) gives the mutation rate in each batch, so we can now calculate the mean probability of mutation across all gametes produced continuously:

\begin{equation} \label{eq:mean}
\begin{split}
	\overline{\mu}&=\dfrac{1}{n_b} \displaystyle \sum_{i=1}^{n_b} \mu_i \\
	&=\dfrac{\mu_d}{n_b} \left( \displaystyle \sum_{i=1}^{n_b-1} \left(\log_2 \left( \dfrac{n_g}{n_b} \right) + i -1 \right)  + \log_2 \left( \dfrac{n_g}{n_b} \right) + n_b -1 -1 \right)\\
	&=\mu_d \times \left( \log_2 \left( \dfrac{n_g}{n_b} \right) -1 + \dfrac{1}{n_b} \left( \displaystyle \sum_{i=1}^{n_b-1} (i) +n_b - 1 \right) \right) \\
	&=\mu_d \times \left( \log_2 \left( \dfrac{n_g}{n_b} \right) -1 + \dfrac{1}{n_b} \left( \dfrac{(n_b-1) n_b}{2} + n_b - 1 \right) \right) \\
	&=\mu_d \times \left( \log_2 \left( \dfrac{n_g}{n_b} \right) -1 + \dfrac{n_b-1}{2} +1 -\dfrac{1}{n_b} \right) \\
	&=\mu_d \times \left( \log_2 \left( \dfrac{n_g}{n_b} \right) + \dfrac{n_b-1}{2} -\dfrac{1}{n_b} \right) 
\end{split}
\end{equation}

Semelparous populations produce a single batch, such that $\overline \mu= \mu_d \times (\log_2(n_g) - 1)$. Interestingly, producing two batches instead of a single one gives the exact same result, because the two batches are produced simultaneously, even though one is going to be used before the other. Once $n_b \geq 3$, the mean mutation rate for gametes produced continuously is above that of gametes produced early. The increase in the mean mutation rate is especially marked when the number of gametes produced, $n_g$, is small, and when the number of batches, $n_b$, is large.

\newpage
\subsection*{Supplementary figures}

\begin{figure}
\centering 
\includegraphics[height=160mm, angle=270]{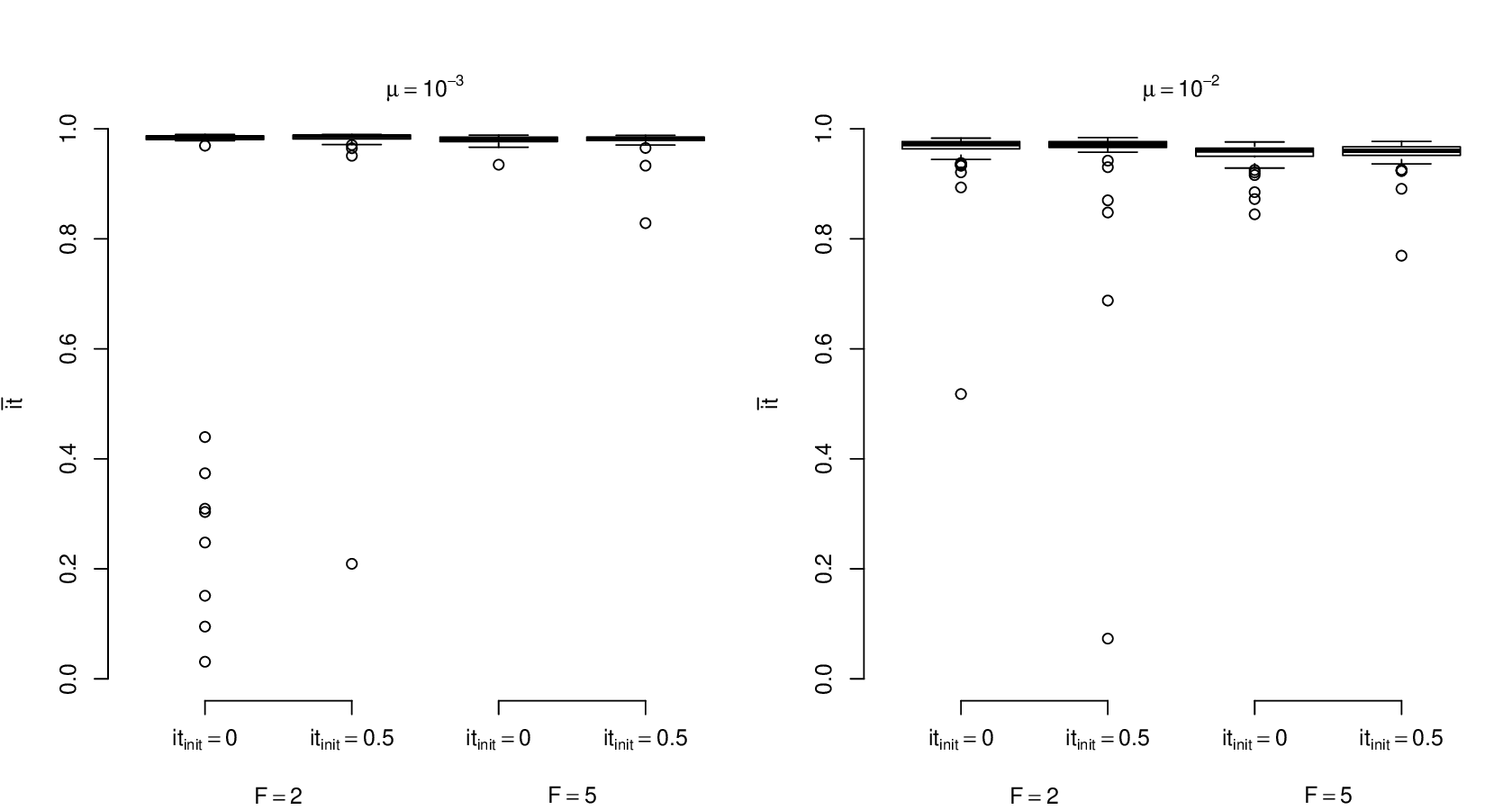}
\caption{Distribution of the mean value of $\alpha$ evolving in $100$ replicate simulations (for each parameter set) where the density dependent egg survival is modeled by function 2, with $K=400$ and $\beta=0.001$. We used two high mutation rates $\mu=10^{-3}$ (left panel) or $\mu=10^{-2}$ (right), with two values for the potential fecundity $F$ ($2$ or $5$). We initiated the simulations with two initial values of $\alpha$: $0$ and $0.5$. }
\label{fig:evol2}
\end{figure}

\begin{figure}[h]
\centerline{\includegraphics[height=87mm, angle=270]{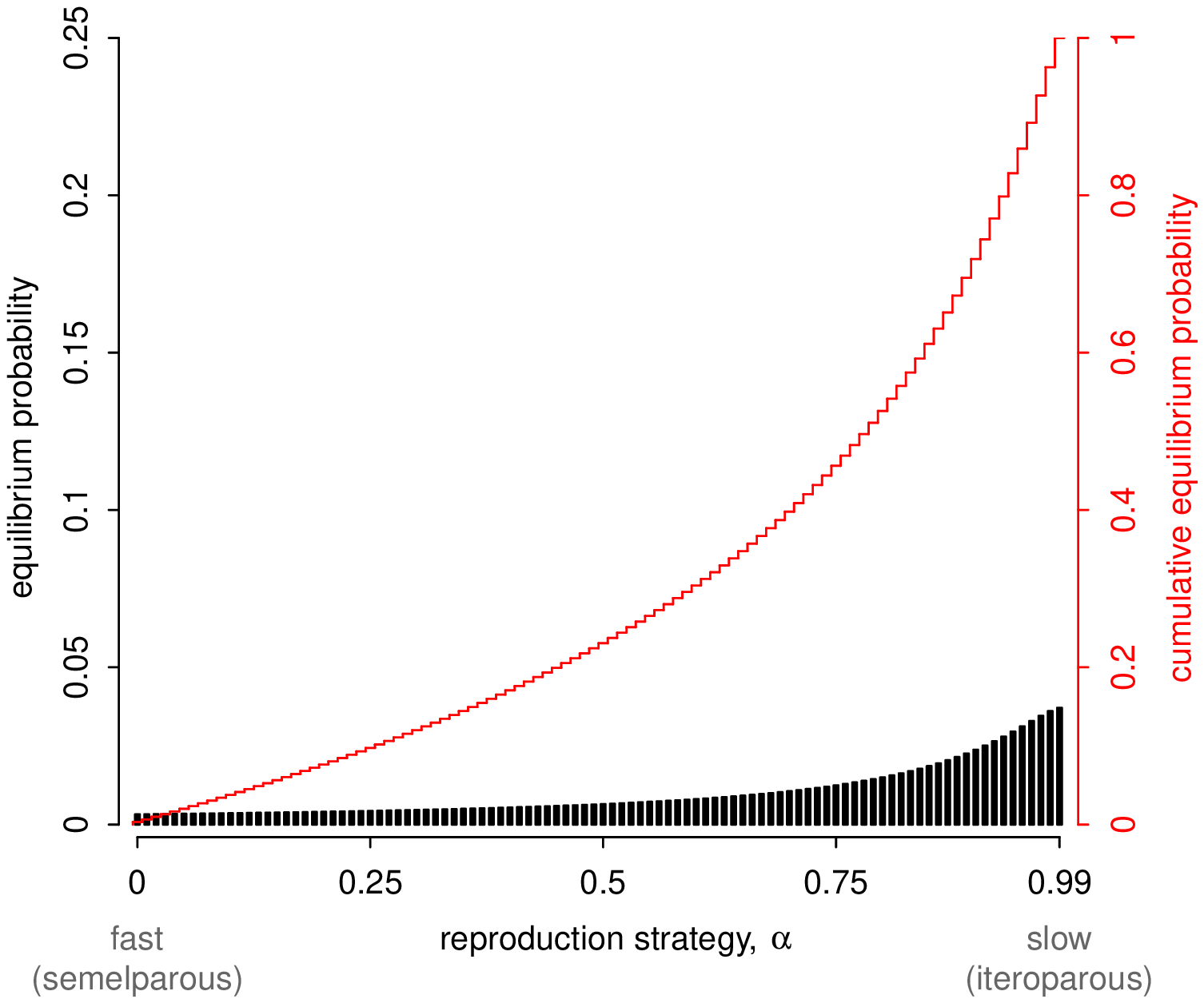}}
\caption{A high degree of iteroparity remains the most likely evolutionary outcome when the mutation rate can increase with an individual's age. The figure is similar to fig.~\ref{fig:mark_res}, except that the model includes the increase in the mutation rate with $\alpha$ (see text). The number of gametes produced in a lifetime, $n_g$, equals $10^2$.}
\label{fig:mark_res3}
\end{figure}

\begin{thebibliography}{10}

\bibitem{takada_1995}
T~Takada.
\newblock Evolution of semelparous and iteroparous perennial plants: Comparison
  between the density-independent and density-dependent dynamics.
\newblock {\em J Theor Biol}, 173:51--60, 1995.

\bibitem{orr_2009}
H~Allen Orr.
\newblock Fitness and its role in evolutionary genetics.
\newblock {\em Nature Reviews Genetics}, 10(8):531--539, 2009.

\bibitem{bulmer_1994}
Michael Bulmer.
\newblock {\em Theoretical evolutionary ecology}.
\newblock Sinauer Associates, 1994.

\bibitem{metz_etal_2008}
J.A.J. Metz, S.D. Mylius, and O.~Diekmann.
\newblock When does evolution optimize?
\newblock {\em Evol Ecol Res}, 10:629--654, 2008.

\bibitem{metz_geritz_2016}
Johan A~J Metz and Stefan A~H Geritz.
\newblock Frequency dependence 3.0: an attempt at codifying the evolutionary
  ecology perspective.
\newblock {\em J Math Biol}, 72(4):1011--37, Mar 2016.

\bibitem{stearns_2000}
S.~C. Stearns.
\newblock Life history evolution: successes, limitations, and prospects.
\newblock {\em Naturwissenschaften}, 87:476--486, 2000.

\bibitem{houle_1998}
D~Houle.
\newblock How should we explain variation in the genetic variance of traits?
\newblock {\em Genetics}, 102:241--253, 1998.

\bibitem{charnov_schaffer_1973}
Eric~L. Charnov and William~M. Schaffer.
\newblock Life-history consequences of natural selection: Cole's result
  revisited.
\newblock {\em Am Nat}, 107(958):791--793, 1973.

\bibitem{schaffer_1974}
W~M Schaffer.
\newblock Optimal reproductive effort in fluctuating environments.
\newblock {\em Am Nat}, 108(964):783--790, 1974.

\bibitem{pianka_1976}
Eric~R Pianka.
\newblock Natural selection of optimal reproductive tactics.
\newblock {\em Amer Zool}, 16(4):775--784, 1976.

\bibitem{bulmer_1985}
MG~Bulmer.
\newblock Selection for iteroparity in a variable environment.
\newblock {\em Am Nat}, pages 63--71, 1985.

\bibitem{kimura_1962}
M.~Kimura.
\newblock On the probability of fixation of mutant genes in a population.
\newblock {\em Genetics}, 47:713, 1962.

\bibitem{metz_etal_1992}
J.A.J. Metz, R.M. Nisbet, and S.A.H. Geritz.
\newblock How should we define 'fitness' for general ecological scenarios?
\newblock {\em Trends Ecol Evol}, 7(6):198--202, 1992.

\bibitem{geritz_etal_1998}
S.A.H. Geritz, E.~Kisdi, G.~Mesz{\'e}na, and J.A.J. Metz.
\newblock Evolutionarily singular strategies and the adaptive growth and
  branching of the evolutionary tree.
\newblock {\em Evol Ecol}, 12(1):35--57, 1998.

\bibitem{proulx_adler_2010}
SR~Proulx and FR~Adler.
\newblock The standard of neutrality: still flapping in the breeze?
\newblock {\em J Evol Biol}, 23(7):1339--1350, 2010.

\bibitem{markovchain}
Giorgio~Alfredo Spedicato.
\newblock {\em markovchain: discrete time Markov chains made easy (R package
  version 0.4.3)}, 06 2014.

\bibitem{R_2013}
{R Core Team}.
\newblock {\em R: A Language and Environment for Statistical Computing}.
\newblock R Foundation for Statistical Computing, Vienna, Austria, 2013.

\bibitem{thomas_etal_2010}
Jessica~A Thomas, John~J Welch, Robert Lanfear, and Lindell Bromham.
\newblock A generation time effect on the rate of molecular evolution in
  invertebrates.
\newblock {\em Mol Biol Evol}, 27(5):1173--1180, 2010.

\bibitem{risch_etal_1987}
N~Risch, EW~Reich, MM~Wishnick, and JG~McCarthy.
\newblock Spontaneous mutation and parental age in humans.
\newblock {\em Am J Hum Genet}, 41(2):218, 1987.

\bibitem{crow_2000}
James~F Crow.
\newblock The origins, patterns and implications of human spontaneous mutation.
\newblock {\em Nat Rev Genet}, 1:40--47, 2000.

\bibitem{petit_hampe_2006}
R\'emy~J. Petit and Arndt Hampe.
\newblock Some evolutionary consequences of being a tree.
\newblock {\em Annu Rev Ecol Evol Syst}, 37:187--214, 2006.

\bibitem{kong_etal_2012}
Augustine Kong, Michael~L Frigge, Gisli Masson, Soren Besenbacher, Patrick
  Sulem, Gisli Magnusson, Sigurjon~A Gudjonsson, Asgeir Sigurdsson, Aslaug
  Jonasdottir, Adalbjorg Jonasdottir, Wendy S~W Wong, Gunnar Sigurdsson,
  G~Bragi Walters, Stacy Steinberg, Hannes Helgason, Gudmar Thorleifsson,
  Daniel~F Gudbjartsson, Agnar Helgason, Olafur~Th Magnusson, Unnur
  Thorsteinsdottir, and Kari Stefansson.
\newblock Rate of de novo mutations and the importance of father's age to
  disease risk.
\newblock {\em Nature}, 488(7412):471--5, Aug 2012.

\bibitem{hurst_ellegren_1998}
L~D Hurst and H~Ellegren.
\newblock Sex biases in the mutation rate.
\newblock {\em Trends Genet}, 14(11):446--52, Nov 1998.

\bibitem{rajon_plotkin_2013}
E~Rajon and J~B Plotkin.
\newblock The evolution of genetic architectures underlying quantitative
  traits.
\newblock {\em Proc. R. Soc. Lond. B}, (280):1769, 2013.

\bibitem{dieckmann_law_1996}
U.~Dieckmann and R.~Law.
\newblock The dynamical theory of coevolution: a derivation from stochastic
  ecological processes.
\newblock {\em J. Math. Biol.}, 34:579--612, 1996.

\bibitem{champagnat_etal_2001}
N.~Champagnat, R.~Ferri{\`e}re, and G~Ben~Arous.
\newblock The canonical equation of adaptive dynamics: a mathematical view.
\newblock {\em Selection}, 2(1--2):73--83, 2001.

\bibitem{mccandlish_stoltzfus_2014}
D~M McCandlish and A~Stoltzfus.
\newblock Modeling evolution using the probability of fixation: history and
  implications.
\newblock {\em Q Rev Biol}, 89(3):225--252, 2014.

\bibitem{proulx_day_2001}
S.R. Proulx and T.~Day.
\newblock What can invasion analyses tell us about evolution under
  stochasticity in finite populations?
\newblock {\em Selection}, 2(1):2--15, 2001.

\bibitem{stoltzfus_2012}
Arlin Stoltzfus.
\newblock Constructive neutral evolution: exploring evolutionary theory's
  curious disconnect.
\newblock {\em Biol Direct}, 7(1):35--35, 2012.

\bibitem{stearns_1983}
S~C Stearns.
\newblock The influence of size and phylogeny on patterns of covariation among
  life-history traits in the mammals.
\newblock {\em Oikos}, 41:173--187, 1983.

\bibitem{gaillard_etal_1989}
J-M Gaillard, D~Pontier, D~Allaine, JD~Lebreton, J~Trouvilliez, and J~Clobert.
\newblock An analysis of demographic tactics in birds and mammals.
\newblock {\em Oikos}, 56(1):59--76, 1989.

\bibitem{Salguero-Gomez_etal_2016}
Roberto Salguero-G{\'o}mez, Owen~R Jones, Eelke Jongejans, Simon~P Blomberg,
  David~J Hodgson, Cyril Mbeau-Ache, Pieter~A Zuidema, Hans de~Kroon, and
  Yvonne~M Buckley.
\newblock Fast-slow continuum and reproductive strategies structure plant
  life-history variation worldwide.
\newblock {\em Proc Natl Acad Sci U S A}, 113(1):230--5, Jan 2016.

\bibitem{braendle_etal_2011}
Christian Braendle, Andreas Heyland, and Thomas Flatt.
\newblock Integrating mechanistic and evolutionary analysis of life history
  variation.
\newblock In Thomas Flatt and Andreas Heyland, editors, {\em Mechanisms of
  life-history evolution -- the genetics and physiology of life history traits
  and trade-offs}, chapter~1. Oxford Univ Press, 2011.

\bibitem{alizon_van-baalen_2005}
Samuel Alizon and Minus van Baalen.
\newblock Emergence of a convex trade-off between transmission and virulence.
\newblock {\em Am Nat}, 165(6):E155--E167, 2005.

\bibitem{cohen_1966}
D.~Cohen.
\newblock Optimizing reproduction in a randomly varying environment.
\newblock {\em J. Theor. Biol.}, 12:119--29, 1966.

\bibitem{seger_brockmann_1987}
J.~Seger and H.~J. Brockmann.
\newblock What is bet-hedging?
\newblock {\em Oxford Surveys in Evolutionary Biology}, 4:182--211, 1987.

\bibitem{orzack_tuljapurkar_1989}
Steven~Hecht Orzack and Shripad Tuljapurkar.
\newblock Population dynamics in variable environments. vii. the demography and
  evolution of iteroparity.
\newblock {\em American Naturalist}, pages 901--923, 1989.

\bibitem{rajon_etal_2009}
E.~Rajon, S.~Venner, and F.~Menu.
\newblock Spatially heterogeneous stochasticity and the adaptive
  diversification of dormancy.
\newblock {\em J. evol. biol.}, 22:2094--2103, 2009.

\bibitem{rajon_etal_2014}
E~Rajon, E~Desouhant, M~Chevalier, F~D{\'e}bias, and F~Menu.
\newblock The evolution of bet hedging in response to local ecological
  conditions.
\newblock {\em The American naturalist}, 184(1):E1--15, 2014.

\bibitem{mitteldorf_martins_2014}
Joshua Mitteldorf and Andr{\'e} C~R Martins.
\newblock Programmed life span in the context of evolvability.
\newblock {\em Am Nat}, 184(3):289--302, Sep 2014.

\bibitem{eyre_keightley_2007}
A.~{Eyre-Walker} and P.~D. Keightley.
\newblock The distribution of fitness effects of new mutations.
\newblock {\em Nat Rev Genet}, 8(8):610--618, 2007.

\bibitem{bulmer_1991}
M.~Bulmer.
\newblock The selection-mutation-drift theory of synonymous codon usage.
\newblock {\em Genetics}, 129(3):897, 1991.

\bibitem{rokyta_etal_2005}
Darin~R Rokyta, Paul Joyce, S~Brian Caudle, and Holly~A Wichman.
\newblock An empirical test of the mutational landscape model of adaptation
  using a single-stranded dna virus.
\newblock {\em Nat Genet}, 37(4):441--4, Apr 2005.

\bibitem{stoltzfus_yampolsky_2009}
Arlin Stoltzfus and Lev~Y Yampolsky.
\newblock Climbing mount probable: mutation as a cause of nonrandomness in
  evolution.
\newblock {\em J Hered}, 100(5):637--47, 2009.

\bibitem{shah_gilchrist_2011}
Premal Shah and Michael~A Gilchrist.
\newblock Explaining complex codon usage patterns with selection for
  translational efficiency, mutation bias, and genetic drift.
\newblock {\em Proc Natl Acad Sci USA}, 108(25):10231--10236, 2011.

\bibitem{xue_etal_2015}
Julian~Z Xue, Andr{\'e} Costopoulos, and Fr{\'e}d{\'e}ric Guichard.
\newblock A trait-based framework for mutation bias as a driver of long-term
  evolutionary trends.
\newblock {\em Complexity}, 2015.

\bibitem{vandooren_metz_1998}
T.~J.~M. van Dooren and J.~A.~J. Metz.
\newblock Delayed maturation in temporally structured populations with
  non-equilibrium dynamics.
\newblock {\em J Evol Biol}, 11:41--62, 1998.

\bibitem{rand_etal_1994}
D.A. Rand, H.B. Wilson, and J.M. McGlade.
\newblock Dynamics and evolution: Evolutionarily stable attractors, invasion
  exponents and phenotype dynamics.
\newblock {\em Phil Trans R Soc B}, 343(1305):261--283, 1994.

\bibitem{ferriere_gatto_1995}
R.~Ferriere and M.~Gatto.
\newblock Lyapunov exponents and the mathematics of invasion in oscillatory or
  chaotic populations.
\newblock {\em Theor Popul Biol}, 48:126--171, 1995.

\bibitem{roff_2008}
Derek~A Roff.
\newblock Defining fitness in evolutionary models.
\newblock {\em J Genet}, 87(4):339--348, 2008.

\bibitem{clermont_1966}
Yves Clermont.
\newblock Renewal of spermatogonia in man.
\newblock {\em Am J Anat}, 118:509--524, 1966.

\end{thebibliography}
\end{document}